\documentclass{jpsj2}
\usepackage{graphicx}
\usepackage{bm}
\usepackage{color}
\def\average#1{\left\langle {#1} \right\rangle}

\def\delo#1{\frac{d}{d #1}}

\def\deld#1#2{\frac{\delta #1}{\delta #2}}
\def\vec2#1#2{\left(\begin{array}{c} #1 \\ #2 \end{array}\right)}
\def\vec3#1#2#3{\left(\begin{array}{c} #1 \\ #2 \\ #3 \end{array}\right)}

\newcommand{\lt}{\left}
\newcommand{\rt}{\right}
\newcommand{\adag}{a^{\dagger}}

\newcommand{\alphaz}{\alpha_{0}}
\newcommand{\Ams}{{\rm A/m}^2}
\newcommand{\Av}{{\bm A}}
\newcommand{\Avem}{{\bm A}_{\rm em}}
\newcommand{\Aph}{A^{\phi}}
\newcommand{\Ath}{A^{\theta}}

\newcommand{\Bv}{{\bm B}}

\newcommand{\Bc}{B_{\rm c}}

\newcommand{\betasf}{{\beta_{\rm sf}}}
\newcommand{\cdag}{c^{\dagger}}
\newcommand{\chiz}{\chi^{(0)}}
\newcommand{\chio}{\chi^{(1)}}
\newcommand{\chitilo}{\tilde{\chi}^{(1)}}
\newcommand{\chitilz}{\tilde{\chi}^{(0)}}
\newcommand{\ckv}{c_{\kv}}

\newcommand{\dx}{{d^3 x}}
\newcommand{\deltan}{\delta n}
\newcommand{\DOS}{{\nu}}

\newcommand{\ef}{{\epsilon_F}}
\newcommand{\eF}{{\epsilon_F}}
\newcommand{\ekv}{\epsilon_{\kv}}
\newcommand{\ekvs}{\epsilon_{\kv\sigma}}
\newcommand{\Ev}{{\bm E}}

\newcommand{\evth}{{\bm e}_{\theta}}
\newcommand{\evph}{{\bm e}_{\phi}}
\newcommand{\evs}{{\bm n}}
\newcommand{\evsz}{{\evs}_0}

\newcommand{\fpin}{f_{\rm pin}}

\newcommand{\Fe}{F}

\newcommand{\Fz}{{F^{(0)}}}

\newcommand{\gr}{g^{\rm r}}
\newcommand{\ga}{g^{\rm a}}
\newcommand{\gless}{g^{<}}

\newcommand{\hf}{\frac{1}{2}}
\newcommand{\HA}{{H_{A}}}

\newcommand{\Hem}{H_{\rm em}}

\newcommand{\Himp}{H_{\rm imp}}

\newcommand{\Hs}{{H_{\rm S}}}
\newcommand{\Hsf}{{H_{\rm sf}}}

\newcommand{\intx}{\int {d^3x}}

\newcommand{\js}{j_{\rm s}}

\newcommand{\jsv}{\bm{j}_{\rm s}}
\newcommand{\jc}{j_{\rm c}}
\newcommand{\jci}{{{j}_{\rm c}^{\rm i}}}

\newcommand{\jatil}{{\tilde{j}_{\rm a}}}
\newcommand{\jctil}{{\tilde{j}_{\rm c}}}

\newcommand{\jtil}{{\tilde{j}}}
\newcommand{\jv}{\bm{j}}
\newcommand{\Jv}{\bm{J}}
\newcommand{\kB}{{k_B}}

\newcommand{\kv}{{\bm k}}

\newcommand{\kf}{{k_F}}
\newcommand{\kfu}{{k_{F+}}}
\newcommand{\kfd}{{k_{F-}}}
\newcommand{\kF}{{k_F}}

\newcommand{\Kp}{{K_\perp}}

\newcommand{\Le}{{L_{\rm e}}}

\newcommand{\Ldw}{{L_{\rm w}}}
\newcommand{\Lp}{L_{\perp}}
\newcommand{\Ls}{{L_{\rm S}}}

\newcommand{\mv}{{\bm m}}

\newcommand{\Mw}{{M_{\rm w}}}

\newcommand{\mub}{\mu_B}
\newcommand{\muB}{\mu_B}
\newcommand{\Ne}{N_{\rm e}}
\newcommand{\nv}{{\bm n}}
\newcommand{\nvz}{{\bm n}_{\rm 0}}
\newcommand{\nimp}{n_{\rm imp}}
\newcommand{\nvortex}{n_{\rm v}}

\newcommand{\Omegatil}{{\tilde{\Omega}}}
\newcommand{\Omegap}{{{\Omega'}}}
\newcommand{\Omegaptil}{{\tilde{\Omegap}}}
\newcommand{\Omz}{{{\Omega_0}}}

\newcommand{\phiz}{{\phi_0}}

\newcommand{\Ptil}{{\tilde{P}}}
\newcommand{\pv}{{\bm p}}
\newcommand{\qv}{{\bm q}}

\newcommand{\renom}{\gamma_S}

\newcommand{\rhos}{{\rho_{\rm s}}}

\newcommand{\Rw}{{R_{\rm w}}}

\newcommand{\Rv}{{\bm R}}
\newcommand{\sigmav}{{\bm \sigma}}
\newcommand{\se}{{s}}
\newcommand{\sev}{{\bm \se}}
\newcommand{\sgn}{{\rm sgn}}

\newcommand{\sv}{{{\bm s}}}
\newcommand{\seth}{{\se}_\theta}
\newcommand{\seph}{{\se}_\phi}
\newcommand{\sez}{{\se}_z}
\newcommand{\spol}{{M}}

\newcommand{\svtil}{\tilde{\bm s}}

\newcommand{\Simpv}{{{\bm S}^{\rm i}}}

\newcommand{\Sv}{{{\bm S}}}

\newcommand{\sumx}{{\int \frac{d^3x}{a^3}}}

\newcommand{\sumqv}{{\sum_{\qv}}}

\newcommand{\thetaz}{{\theta_0}}
\newcommand{\tr}{{\rm tr}}
\newcommand{\Tc}{{T_{C}}}

\newcommand{\torque}{{\tau}}
\newcommand{\torquev}{{\bm \torque}}

\newcommand{\torqueve}{{\torquev_{\rm e}}}
\newcommand{\torquee}{{\torque_{\rm e}}}

\newcommand{\tautil}{{\tilde{\tau}}}

\newcommand{\ttil}{{\tilde{t}}}

\newcommand{\vc}{{v^{\rm c}}}

\newcommand{\vf}{{v_F}}
\newcommand{\vimp}{v_{\rm imp}}

\newcommand{\Vpin}{{V}_{\rm pin}}

\newcommand{\Vz}{{V_0}}
\newcommand{\Vztil}{{\tilde{V_0}}}
\newcommand{\Ws}{{W_{\rm S}}}
\newcommand{\Xtil}{{\tilde{X}}}
\newcommand{\xv}{{\bm x}}

\newcommand{\xw}{{z}}

\title{
Theory of Domain Wall Dynamics under Current}
\author{
Gen Tatara$^{1,2}$,
Hiroshi Kohno$^{3}$ and
Junya Shibata$^{4}$
}

\inst{
$^1$Graduate School of Science and Engineering, Tokyo Metropolitan University\\
1-1 Minamiosawa, Hachioji,  Tokyo 192-0397\\
$^2$PRESTO, JST, 4-1-8 Honcho, Kawaguchi, Saitama 332-0012\\
$^3$
Graduate School of Engineering Science, Osaka University,
Toyonaka, Osaka 560-8531\\
$^4$
Kanagawa Institute of Technology,
1030 Shimo-Ogino, Atsugi, 243-0292
\\
}

\abst{
Microscopic theory of domain wall dynamics under electric current is reviewed. 
Domain wall is treated as rigid and planar.
The spin-transfer torque and forces on the wall are derived based on the $s$-$d$ exchange interaction between localized spins and conduction electrons, treating non-adiabaticity expressed by the gauge field perturbatively. 
Effect of spin relaxation is also studied. 
}

\kword{
spin torque, domain wall, magnetoresistance,  
Berry phase, MRAM
}

\begin{document}
\maketitle
\section{Introduction}

The present information technology is based on electron transport and magnetism.
Magnetism has been most successful as high-density storages such as hard disks.
In magnetic storages, read-out mechanism of the information had so far several significant developments.
The oldest idea of detecting magnetic information would be to use Faraday's induction law by scanning a read head (a coil) on the stored magnetic bits.
More efficient read-out mechanism was developed by using anisotropic magnetoresistance (AMR) effect. 
AMR is a resistivity dependent on the angle between the magnetization and the electric current, which arises from the coupling of magnetization and electrons' oribtal motion due to spin-orbit interaction\cite{McGuire75}.
The resistivity change is only of order of a few \%, but AMR is more efficient than using Faraday's induction used in magnetic tape and hard disks in early days.
Magnetic head with higher sensitivity was developed by use of GMR (giant magnetoresistance) effect in thin magnetic multilayers discovered in 1988\cite{Baibich88,Binasch89}. 
Strong magnetization dependence of the resistivity arises from the spin-dependent scattering of the electron at the interface between thin ferromagnetic layer and non-magnetic metallic layers. 
Quite recently GMR heads are being replaced by TMR (Tunneling MR) heads, where the non-magnetic layer is replaced by an insulating barrier\cite{Miyazaki95}.
These rapid developments of read-out mechanism by use of solid state systems made possible so far the rapid increase of recording density. 

\subsection{Current-driven domain wall motion}

In contrast, write-in mechanism in magnetic devices is still based on the knowledge of 19th century, Amp\`ere's law.
A novel mechanism of controlling magnetization by use of current but without referring to magnetic field was first considered by Berger in 1978\cite{Berger78}.
He pointed out theoretically a possibility of driving a domain wall by current directly.
Domain wall is a twisted structure of local spins as shown in Fig. 1 
\cite{Hubert98,Marrows05}.

\begin{figure}[tbh]
\label{DW}
\begin{center}
\includegraphics[width=0.4\linewidth]{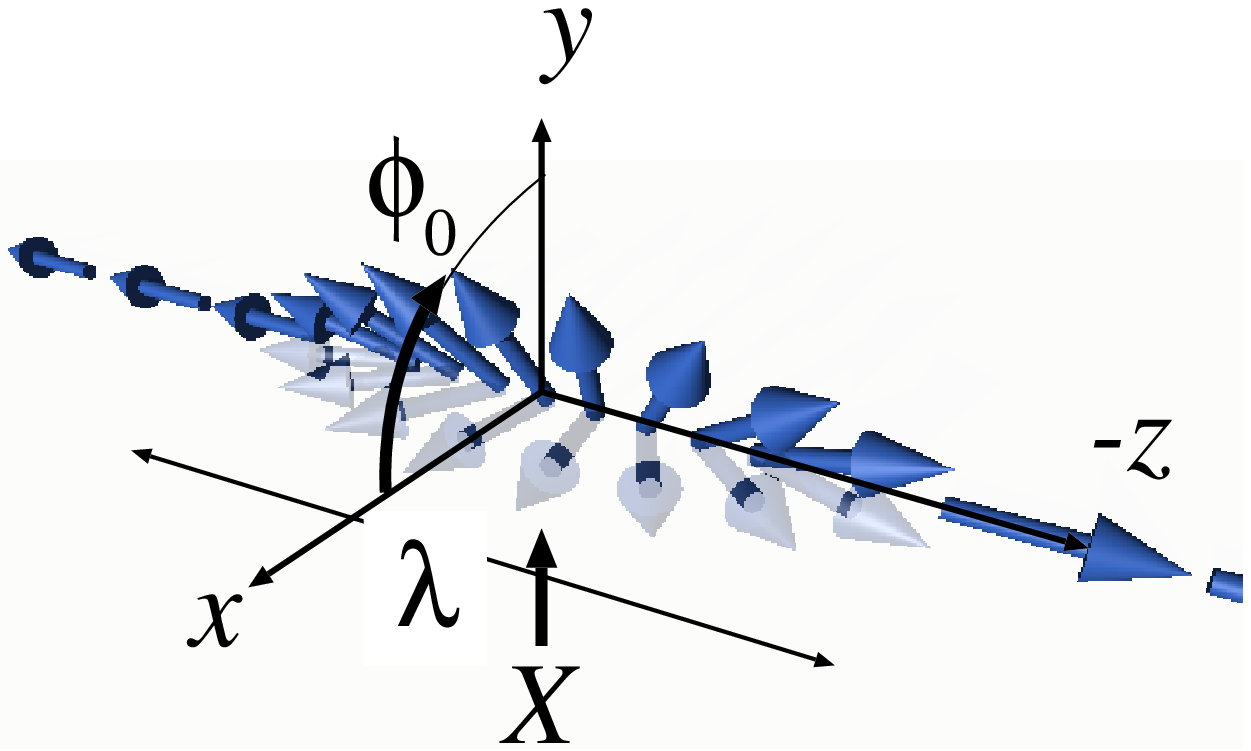}
\includegraphics[width=0.4\linewidth]{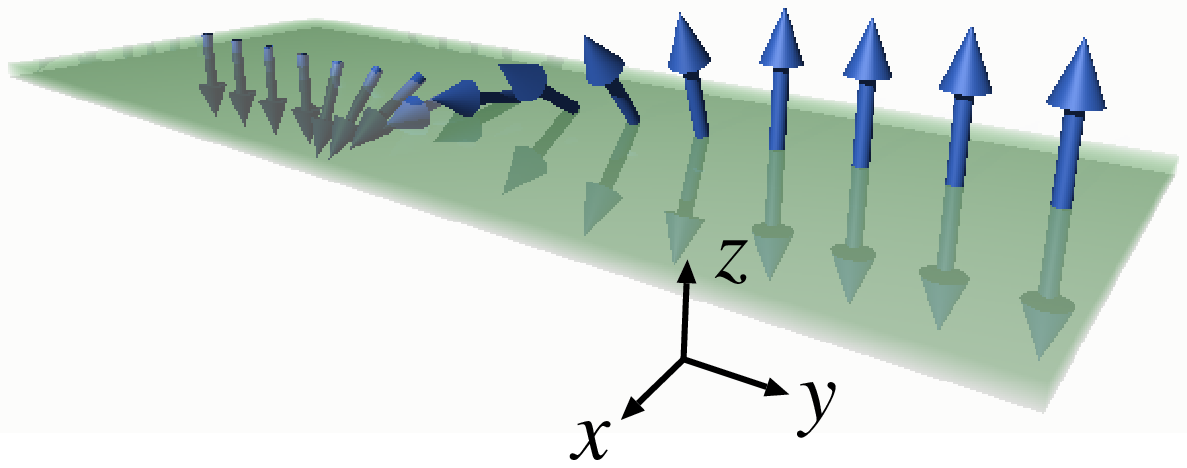}
\label{FIGdw}
\caption{ (color online)
 Illustration of a N\'eel wall (left) and a Bloch wall (right). 
 Also shown are the wall position $X$ and the angle $\phiz$ between the wall magnetization and the easy plane, which are collective coordinates describing a rigid and one-dimensional (or planar) wall. $\lambda$ is the thickness of the wall. 
The axes shown are those in spin space (they do not necessarily coincide with spatial coordinate axes in the present systems without coupling between spin and real spaces). 
The current direction is perpendicular to the wall plane.
}
\end{center}
\end{figure} 

The thickness of the wall, $\lambda$, is an important parameter governing the coupling between the wall and conduction electrons.
In most cases of 3d transition metals, 
$\lambda$ is $10-100$nm, and is much larger than the Fermi wavelength of the electron, $\kF^{-1}$.
The wall is therefore in the adiabatic limit, where the elctron spin can adiabatically follow the local spin direction as it passes through the wall.
We consider in this paper a domain wall which is planar and rigid. In this case, the wall dynamics is described 
by two variables, its position $X$ and $\phiz$, average angle out of easy plane ($x$-$z$ plane in both walls in Fig. 1)\cite{Slonczewski72,TK04}.
Our calculation applies to both N\'eel and Bloch walls, since there is no coupling between spin and real space in our model. 
The current direction is perpendicular to the wall plane.

Berger considered a domain wall under electric current, and discussed that $s$-$d$ type exchange coupling between the local spin and conduction electron spin is dominant interaction that drives the wall under current in the case of thin film 
(e.g.,  thickness less than $\sim 0.1\mu$m), 
where the effect of induced magnetic field can be neglected\cite{Berger78}.
In 1984\cite{Berger84,Hung88}, he studied the effect of the force arising from the reflection of conduction electron by domain wall caused by this exchange coupling. 
This force was associated with a wall mobility introduced phenomenologically.
The effect was found to be small in most cases due to a very small reflection probability because the wall thickness is usually large compared with Fermi wavelength.
In 1986\cite{Berger86}, he argued that the exchange interaction produces a torque which tends to cant the wall out of the easy plane (angle $\phiz$).
This torque was later found to push the wall by different mechanism from exchange force, and turned out to be dominant driving mechanism\cite{Berger92}, and is nowadays called spin transfer torque.
Based on this idea, an experimental study was carried out in 1993\cite{Salhi93} on a thin film of Ni$_{81}$Fe$_{19}$. 
There, domain wall velocity of 70 m/s
was reported at the current density of 
$1.35\times 10^{10}$ A/m$^2$ applied as a pulse of duration 0.14$\mu$s.
Although the experiment was quite successful, experiments on better-controlled systems are now required.

After these works, there was no significant development until 1996 in studies on current-driven domain wall motion, when Slonczewski\cite{Slonczewski96} and Berger\cite{Berger96} independently developed a theory of magnetization switching by spin-transfer torque in thin film or pilar structures.
This spin-transfer torque is essentially the same as the one Berger has discussed for domain wall\cite{Berger92}.
The pillar system considered there was intensively studied after the works by Slonczewski and Berger since such a system is expected to be applied to a memory devices like magnetoresistive ramdom access memory (MRAM) that opeartes without magnetic field.
Current-induced domain wall motion is also expected to be useful as a possible MRAM, and intensive experimental studies have started.

Recently experimental studies have been carried out on submicron-size wires and domain wall motion induced by current has been confirmed\cite{Grollier02,Grollier03,Tsoi03,Klaui03}.
In all of the early experiments, the current density necessary for wall motion is 
high, of order of $10^{11}-10^{12}$ A/m$^2$. 
Measurement of domain wall velocity was carried out by Yamaguchi et al. \cite{Yamaguchi04} by observing wall displacement by use of MFM after each current pulse of strength $1.2\times 10^{12}$ A/m$^2$ and duration of 5$\mu$s.
The average velocity was found to be $2\sim 6$m/s.
In those experimental works, there seemed to be a certain threshold
value for domain wall motion, around $10^{12}$ A/m$^2$, and the average wall velocity were rather slow, less than 10m/s.

\subsection{Recent theories}

Those experiments motivated theoretical studies to look into the problem in more detail. 
Microscopic derivation of equation of motion of domain wall under current was carried out by Tatara and Kohno\cite{TK04}.
They considered a planar (one-dimensional) wall and described the wall by the two collective coordinates, $X$ and $\phiz$, i.e. within Slonczewski's description\cite{Slonczewski72}. 
The variable $X$ represents the position of the wall, and $\phiz$ describes a tilt of the wall plane.  
Considering a small hard-axis anisotropy case, other deformation modes than $\phiz$ (such as change of wall width) were neglected (rigid wall approximation).  
The equation of motion with respect to $X$ and $\phiz$ was derived including the effect of conduction electrons via the $s$-$d$ exchange interaction.
The electron carrying a current was treated by use of non-equillibrium (Keldysh) Green function.
The equation of motion derived was found to be essentially the same as that obtained by Berger long ago\cite{Berger84,Berger92}, indicating his deep physical insights, but the effects of the spin-transfer torque and reflection force (momentum transfer) were obtained without phenomenological assumptions and ambiguities for the first time. 
Based on the obtained equation of motion, the wall motion under steady current was studied. 
It was found that in the adiabatic limit, where the reflection force can be neglected, and if in the absence of spin relaxation, 
there is a threshold current determined by the hard-axis magnetic anisotropy energy, $\Kp$.
Thus the wall is intrinsically pinned by the internal degree of freedom, $\phiz$.
At large current, however, the wall gets depinned and its velocity becomes proportional to spin current (spin polarization of the current flow), $\js$, as is required from the angular momentum conservation.

Numerical simulation was performed based on an equation of motion of each local spin
by including the spin-transfer torque term in the adiabatic limit \cite{Thiaville04}. 
The result was similar to the analytical (collective-coordinate) study, indicating the existence of threshold current. 
Motion of domain wall under magnetic field and spin-transfer troque was solved in Ref. \cite{Li04}.
Later Zhang and Li \cite{Zhang04} and Thiaville et al.\cite{Thiaville05} proposed to add a new torque term in the equation, which is perpendicular to the spin-transfer torque. 
 After Thiaville et al.\cite{Thiaville05}, 
we call this torque term as the $\beta$ term.
Zhang and Li argued that the $\beta$ term 
arises from spin relaxation of conduction electrons\cite{Zhang04}. 
Thus the phenomenological equation of motion of local spin under current reads
\begin{equation}
\dot{\Sv}= \Bv_{\rm eff} \times \Sv
+\frac{\alpha}{S} \Sv\times\dot\Sv
-\frac{a^3}{2eS}(\jsv\cdot\nabla)\Sv
- \frac{a^3 \beta}{eS} [\Sv \times (\jsv\cdot\nabla)\Sv]
+{\bm \tau}_{\rm na}.
\label{modLLG}
\end{equation}
Here $\Bv_{\rm eff}$ is the effective field arising from spin Hamiltonian, and $\alpha$ represents damping.
The equation $\dot{\Sv}= \Bv_{\rm eff} \times \Sv
+\frac{\alpha}{S} \Sv\times\dot\Sv$ has been well-known as Landau-Lifshitz-Gilbert equation describing magntization dynamics in a magnetic field $\Bv_{\rm eff}$. 
Effects of current are represented by other three terms in Eq. (\ref{modLLG}).
The third term on the right hand side, $(\jsv\cdot\nabla)\Sv$, represents spin transfer torque, 
the fourth one is $\beta$ term, 
and the last term ${\bm \tau}_{\rm na}$ denotes non-adiabatic term\cite{TKSLK07}, which has not been taken account in numerical simulations.

 The $\beta$ term turned out to modify the threshold current and the terminal velocity of the wall significantly when $\beta/\alpha\gtrsim1$\cite{Zhang04,Thiaville05,TTKSNF06}.
The threshold current in this case is determined by extrinsic pinning and the terminal wall velocity is determined by $\beta/\alpha$
\cite{Zhang04,Thiaville05,TTKSNF06}.
Microscopic derivation of the $\beta$-term was carried out by several authors
\cite{Tserkovnyak06,KTS06,KS07,Duine07}. 
Tserkovnyak et al.\cite{Tserkovnyak06} calculated $\beta$ based on a one-band model considering spin-relaxation of conduction electrons semi-classically and assuming spin dynamics of small amplitude. 
They considered the limit of weak ferromagneticm and found that 
$\beta_{\rm sf}=\alpha_{\rm sf}$, namely, $\beta$ due to spin flip is equal to the damping parameter caused by spin flip.
They also mentioned that in general 
$\beta_{\rm sf} \neq \alpha_{\rm sf}$ considering the effects of multiband or deviation from weak ferromagnetism.  
Their approach is, however, still phenomenological, treating the spin-flip process by a phenomenological spin-relaxation time in the equation of motion of spin. 
$\beta_{\rm sf}=\alpha_{\rm sf}$ was suggested also by different phenomenological argument\cite{Barnes05,Barnes06} (see also Ref. \cite{TK06}).
Fully microscopic calculation of $\beta$ and $\alpha$ due to spin relaxation was carried out on s-d model by Kohno et al.\cite{KTS06,KS07} using  standard diagrammatic perturbation theory, where effect of spin relaxation are taken into account consistently and fully quantum mechanically. 
The result indicated $\beta_{\rm sf}\neq \alpha_{\rm sf}$.
The same result was obtained later in the functional Keldysh formalism by Duine et al.\cite{Duine07}.
 Determination of $\beta$ and $\alpha$ values needs a careful microscopic calculation, since they are quantities smaller by a factor of $1/(\ef\tau)$ compared with conventional transport coefficients. 
Phenomenological thermodynamic argument predicted $\beta=\alpha$ \cite{Barnes05,Barnes06}, but microscopic studies\cite{KTS06,Duine07,KS07} indicate that it is wrong. 
The error would be because the argument of refs. \cite{Barnes05,Barnes06} lacks consistent consideration of the work done by the electric current\cite{TK06}.

Waintal and Viret\cite{Waintal04} and  Xiao, Zangwill and Stiles\cite{Xiao06} studied the spatial distribution of the current-induced torque around a domain wall by solving Schr\"odinger equation and found a non-local oscillatory torque
($\tau_{\rm na}$ in eq. (\ref{modLLG})). 
This torque is due to the non-adiabaticity arising from the finite domain wall width, or in other words,
from the fast-varying component of spin texture. 
The oscillation period is $\sim \kf^{-1}$ ($\kf$ is the Fermi wavelength) and is of quantum origin similar to the RKKY oscillation.
Ohe and Kramer\cite{Ohe06} studied a wall motion solving the torque due to the exchange interaction numerically, including
non-adiabaticity. 
Non-local oscillating torque was numerically studied by taking account of strong spin-orbit interaction based on Kohn-Luttinger Hamiltonian  (i.e., in magnetic semiconductors)\cite{Nguyen07}.
It was shown there that the oscillating torque is asymmetric around domain wall and that this feature results in high wall 
velocity.
Non-adiabaticity was studied further in Refs. \cite{TKSLK07,Piechon06,Thorwart07}.

In this paper, we review recent developments in the theory of current-driven domain wall motion.
We consider the case of a rigid one-dimensional (planar) domain wall. 
This rather drastic assumption turns out to be more or less valid when compared with the numerical simulation and some of the experimental data available at present. 
Detailed comparison of experimental results with the present study will also shed light on the role of deformation of the wall, which is the future target.

\section{Model}

 For simplicity, we take a localized picture for ferromagnetism and adopt the $s$-$d$ model, where the dominant part of the ferromagnetic moment is carried by localized $d$-spins, $\Sv(\xv,t)$, and they are coupled to conduction electrons via the $s$-$d$ exchange interaction. 
 Essentially the same description would hold for itinerant ferromagnets, where the ferromagnetic order parameter plays the role of $\Sv(\xv,t)$ above.\cite{TF94}

The Lagrangian of the system consists of that of electrons $\Le$, that of localized spins $\Ls$, and the $s$-$d$ exchange interaction $H_{sd}$ between them;  
\begin{equation}
 L = \Le + \Ls - H_{sd}.
\end{equation}
 Each term will be explained in the following. 
 Starting from this Lagrangian, we will derive the equation of motion of a domain wall. 
 Since the domain wall is a macroscopic object, we treat it classically, whereas conduction elctrons are treated quantum mechanically.

\subsection{Electron part}

The electrons we consider are interacting with impurities (both non-magnetic and magnetic) and with external eletric field. 
We denote the electron annihilation and creation operators as $c_{\sigma}(\xv,t)$ and $\cdag_{\sigma}(\xv,t)$,  respectively, where $\sigma=\pm$ represents the spin state.
The total electron Lagrangian is given by
\begin{align}
\Le &=   \sum_{\kv} c^\dagger_{\kv} \lt(i\hbar \partial_t-\ekv\rt)  \ckv
- \Himp - \Hsf -\Hem
,\label{LE}
\end{align}
where
$\ekv\equiv \frac{\hbar^s k^2}{2m}-\eF$ and 
$\ckv\equiv (c_{\kv+},c_{\kv-})^{\rm t}$ (t denotes transposition).
The spin-independent impurity scattering is described by
\begin{align}
\Himp
 &= \sum_{i}\sum_{\kv,\qv} 
 \vimp e^{-i\qv \cdot\Rv_i} c^\dagger_{\kv+\qv}c_{\kv},
\end{align}
where $\vimp$ represents the potential due to impurity and $\Rv_i$ represents position of random impurities.
We approximate the potential as on-site, i.e., $\vimp$ has no $\qv$-dependence.
Treating the impurity scattering  by Born approximation, $\Himp$ gives rise to a lifetime of the electron, $\tau_\sigma$, whose inverse is given by
\begin{equation}
 \frac{1}{\tau_\sigma} = 2\pi\nimp\vimp^2 \DOS_\sigma,
\end{equation}
where $\DOS_\sigma$ is the density of states at the Fermi level and $\nimp$ is the concentration of impurities.
The density of states and lifetime are spin-dependent in general, but we neglect this spin dependence when we discuss the spin transfer and momentum transfer, to avoid unnecessary complexity. 
This spin-dependence becomes important and so will be retained later when the effect of spin relaxation is studied.

The term $\Hsf$ represents spin-flip scattering due to random magnetic impurities:
\begin{equation}
\Hsf = u_s \intx \sum_i \Simpv_i \delta(\xv-\Rv'_i)
(\cdag\sigmav c)_\xv,
\end{equation}
where $\Simpv_i$ represents impurity spin at site $\Rv'_i$.
A quenched average is taken for the impurity spin direction as well as for the impurity position.

The interaction with electric field is expressed by use of charge current $\jv$ and electromagnetic gauge field $\Avem$ as
\begin{equation}
\Hem = - \int \dx \, \Avem \cdot \jv,
\end{equation}
The gauge field is given by use of $\Ev\equiv - \dot{\Av}_{\rm em}$ as 
\begin{equation}
\Avem= i\frac{\Ev}{\Omz}e^{i\Omz t},
\end{equation}
where $\Ev$ is the applied electric field which is spatially homogeneous, and $\Omz$ is its frequency to be set as $\Omz\rightarrow0$ at the last stage of the calculation. 
The total current is given in the presence of gauge field by
($e(<0)$ is electron charge)
\begin{equation}
\Jv \equiv  \frac{e}{m} 
\sum_{\kv} (\hbar \kv - e \Avem ) 
c^{\dagger}_{\kv}c_{\kv} .
\label{current}
\end{equation}

\begin{figure}[tbh]
  \begin{center}
  \includegraphics[width=0.3\linewidth]{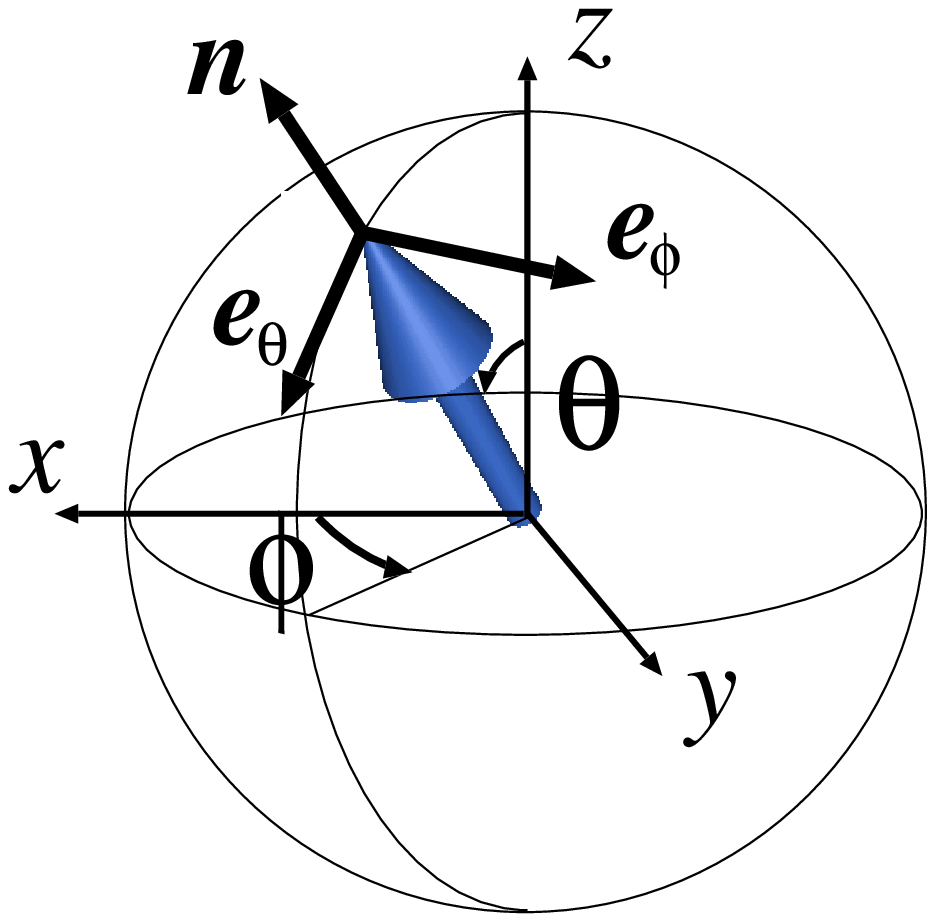}
  \end{center}
\label{FIGpolar2}
\caption{Polar coordinates $(\theta,\phi)$ parametrizing the local spin direction, and unit vectors $\evs$, $\evth$ and $\evph$.
}
\end{figure}

\subsection{Spin part}

We consider the magnetization, or local spins, of fixed magnitude $S$ and 
varying direction ${\bm n}$, and parametrize it by the polar coordinates $(\theta,\phi)$ as (Fig. \ref{FIGpolar2}) 
\begin{equation}
\Sv=S (\sin\theta\cos\phi,\sin\theta\sin\phi,\cos\theta)
\equiv S \evs.
\end{equation}
Deferring the effect of damping (friction) to the next subsection, 
the spin part of the Lagrangian is given by
\begin{equation}
\Ls = \hbar S \sumx \dot{\phi}(\cos\theta -1) -\Hs,
\label{Ls}
\end{equation}
where $\Hs$ is the Hamiltonian of local spin, which we will specify later. 
The first term  is known as the \lq kinetic potential', and describes the spin dynamics governed by a torque equation. 
It has the same form as the spin Berry phase in quantum mechanics, but here we treat localized spins as classical objects. 
 In fact, the equation of motion is derived from $\Ls$
as
\begin{equation}
 \dot{\Sv} = \gamma\Bv_{\rm s}\times\Sv, \label{torqueeq}
\end{equation}
 where 
$\gamma\Bv_{\rm s}\equiv \deld{\Hs}{\Sv}$ is the effective magnetic field acting on localized spin (in the absence of conduction electrons).

The meaning of the \lq spin Berry phase' term can be understood if one note that the canonical structure is contained in this kinematical term in the Lagrangian. 
Let us demonstrate this within classical mechanics.
The canonical momentum conjugate to $\phi$ is defined as 
\begin{equation}
 P_\phi \equiv \deld{\Ls}{\dot{\phi}} 
 = \hbar S_z -\hbar S .
\end{equation}
 Defining the Poisson bracket (times $\hbar$) by 
$\{ A , B \}_{\rm PB} 
= (\partial A/\partial \phi) (\partial B/\partial S_z) 
- (\partial B/\partial \phi) (\partial A/\partial S_z)$, 
we have $\{ \phi , S_z \}_{\rm PB} = 1$. 
 By using 
$S_x \pm iS_y = \sqrt{S^2 - S_z^2} \, e^{\pm i\phi}$,
we can derive the correct SU(2) algebra of the spin angular momentum as 
\begin{align}
 \{ S_i,S_j \}_{\rm PB} &= \epsilon_{ijk} S_k .
\label{spin:Scom}
\end{align}

The Hamiltonian of localized spin we consider is a general one with two anisotropy energies.
The easy axis  and a hard axis, chosen as $z$ and $y$ direction, respectively. 
We treat local spins in the continuum.
The Hamiltonian is given by
\begin{align}
\Hs &= \sumx \lt[ \frac{J}{2}(\nabla \Sv)^2 - \frac{K}{2} (S_z)^2  
     + \frac{\Kp}{2} (S_y)^2 \right] +\Vpin,
  \label{spin:Hs}
\end{align}
where $\Vpin$ represents 
sample inhomogeneity leading to the pinning of a domain wall.
 For a wire of soft ferromagnet, the easy axis is in the wire direction to avoid surface magnetic charges (shape anisotropy), and a domain wall appears as the N\'eel wall (Fig. \ref{FIGdw}).
The case of Bloch wall such as realized in a film with perpendicular magneic anisotropy is also described by $\Hs$.
As for the spin-transfer and momentum-transfer processes, both types of domain walls show the same dynamics if the spin-orbit coupling is neglected in the electron system.

\subsection{Damping}

In spin dynamics, damping (friction) plays an essential role. 
 We know that the magnetization will eventually point to the direction of the effective field, $\Bv_{\rm s}$. 
Simple torque equation, $\dot\Sv = \gamma \Bv_{\rm s}\times \Sv$ ($\gamma\equiv \frac{|e|}{m}$ is gyromagnetic ratio), or 
 $\dot\nv = \gamma \Bv_{\rm s}\times \nv$, however,  predicts only a precession around $\Bv_{\rm s}$.
This point was remedied by Landau and Lifshitz (LL) by adding a perpendicular torque, 
\begin{equation}
 \dot{\nv} = \gamma \Bv_{\rm s}\times\nv
 + \alpha_{\rm LL} (\nv\times(\nv\times \Bv_{\rm s})), \label{LL}
\end{equation}
where the last term describes a damping torque, which tends to align $\nv$ along $\Bv_{\rm s}$.
Gilbert later proposed another form of damping, which contains $\dot{\nv}$, 
\begin{equation}
 \dot{\nv} = \gamma \Bv_{\rm s}\times\nv
 - \alphaz (\nv\times \dot{\nv}), \label{LLG}
\end{equation}
and this equation (\ref{LLG}) is called Landau-Lifshitz-Gilbert (LLG) equation. 
(In the above, $\alpha_{\rm LL}$ and $\alphaz$ are dimensionless parameters.)
These two equations are essentially equivalent, and they describe correctly the decay of precession and the relaxation to the equillibrium direction, $\nv\parallel \Bv$.
As we will see, damping terms of Gilbert form can be derived by integrating out the environment. 
(But damping torque has also higher order derivatives, and so both LL and LLG equations are approximations in the linear order of time derivative.)

The damping term cannot be introduced in the Lagrangian 
without environment, as is always the case for dissipation processes. 
We here treat the Gilbert damping by use of Rayleigh's method in classical mechanics\cite{Goldstein02}, by considering a quantity describing energy dissipation, 
\begin{equation}
\Ws\equiv \frac{\alphaz}{2} \hbar S  \sumx 
\,(\dot\theta^2 + \sin^2\theta \, \dot\phi^2) .
\end{equation}
The equation of motion with damping included is given by 
\begin{equation}
\delo{t} \deld{\Ls}{\dot{q}}
-\deld{\Ls}{q} = -\deld{\Ws}{\dot{q}} , 
\label{spineq}
\end{equation}
where $q$ represents $\theta$ and $\phi$.
The derived equation is the LLG equation, (\ref{LLG}).

\subsection{Exchange interaction}

The most important interaction in our problem is the $s$-$d$ exchange interaction 
\begin{equation}
H_{sd} = - J_{sd} \!\intx \, 
  \Sv \cdot (c^\dagger \sigmav c) 
\equiv - M \!\intx \, 
  \nv \cdot (c^\dagger \sigmav c) , 
\end{equation}
between local spin and conduction electrons. 
 Here $M \equiv J_{sd}S$ is half the exchange splitting. 
 An important point is that $J_{sd}$ is rather strong in 3d ferromagnets: 
$J_{sd}/\eF \gtrsim O(0.1 -1)$. 
 These values are indicated from experimental observations of large magnetoresistances such as GMR.

Most non-trivial part of the theory is the treatment of this strong exchange interaction when local spin has a spatial structure and/or is dynamical. 
 Fortunately, spin structures in 3d ferromagnets are slowly varying compared to the scale of conduction electrons. 
 This is a conseqence of strong exchange interaction, $J$, between local spins, which is of order of 1000K as indicated by high critical temperature of 3d ferromagnets (For Fe, $\Tc\sim 1043$K). 
(Correctly, the typical length scale, $\lambda$, is determined by the ratio of exchange energy and magnetic anisotropy 
(Eq. (\ref{thickness})).) 
Since many local spins within the scale of $\lambda$ are coupled, the spin structure is (semi-) macroscopic and its time scale is slow compared to that of electrons. 
From these considerations, the electron can go through the spin structures adiabatically. 
The condition for the adiabaticity can be given by a few different small parameters.
The first one, introduced by Stern\cite{Stern92} in disordered case,
\begin{equation}
\hbar/(\spol \tau) \ll 1, \label{adiabatic1}
\end{equation}
justifies the perturbative treatment of non-adiabaticity by using a gauge field.
The second small parameter expressing spatially slow variation is $1/(\kF \lambda) \ll1$.
For spin transport in the absence of disorder, this condition  would be modified to be 
\begin{equation}
\frac{1}{\kF \lambda}\frac{\eF}{\spol} \simeq \frac{1}{(\kfu-\kfd)\lambda} \ll 1 , \label{adiabatic2}
\end{equation}
where the left hand side is a ratio of the precession time of conduction electron due to the exchange interaction, $\hbar/\spol$, to the time needed for the electron to pass through the spin structure,  $\lambda/\vf$, as proposed by Waintal and Viret\cite{Waintal04}.

\section{Gauge Transformation}

\begin{figure}[tbh]
\label{FIGgaugetr}
\begin{center}
\includegraphics[width=0.3\linewidth]{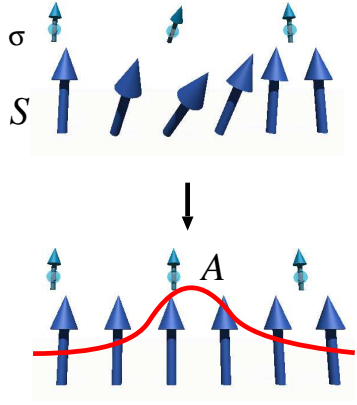}
\caption{Gauge transformation of electrons in inhomogeneous field (due to local spins, $\Sv$) into electrons in homogeneous field produces an SU(2) gauge field, $\Av$.
}
\end{center}
\end{figure}

Under the condition of Eq.(\ref{adiabatic1}), the electron spin is polarized almost along the local spin direction. 
 By use of local gauge transformation in spin space, such electrons are mapped to electrons in a uniform ferromagnetic state interacting with a gauge field. 
The local gauge transformation is to choose the electron spin quantization axis along $\Sv(\xv,t)$ at each point (Fig. \ref{FIGgaugetr})\cite{TF94}.
The deviation from perfect adiabaticity is described by an SU(2) gauge field, which is small and we treat it perturbatively.
A new electron operator $a\equiv ({a_+},{a_-})^{\rm t}$ 
is defined as
\begin{equation}
c(\xv,t)\equiv U(\xv,t) a(\xv,t),
\end{equation}
where $U$ is a $2\times2$ matrix which we take here as
\begin{equation}
U(\xv,t)\equiv \mv\cdot\sigmav,
\end{equation}
with a unit vector $\mv$ given by 
\begin{equation}
\mv=\left(
\sin\frac{\theta}{2}\cos\phi,\sin\frac{\theta}{2}\sin\phi,\cos\frac{\theta}{2} \right).
\end{equation}
 The derivative of the original electron is written as 
$ \partial_\mu c = U (\partial_\mu+iA_\mu) a $
in terms of new electron, where 
$ A_\mu \equiv -iU(\xv,t)^{-1}\partial_\mu U(\xv,t) 
        \equiv A_\mu^\alpha \sigma_\alpha $ 
is the SU(2) gauge field with 
$A_\mu^\alpha \equiv (\mv\times\partial_\mu \mv)^\alpha$ 
(summation over $\alpha=x,y,z$ is suppressed).
 The gauge field is related to the derivative  $\evs$, as 
$ \partial_\mu \evs=2 \Av_\mu\times \evs$.

The free-electron part of the Lagrangian 
is written in terms of the $a$-electron as 
\begin{align}
\sum_{\kv,\sigma} c^\dagger_{\kv\sigma}
( i\hbar\partial_t -\epsilon_{\kv} ) c_{\kv\sigma} 
&=  
\sum_{\kv,\sigma}  a^\dagger_{\kv\sigma}
( i\hbar\partial_t -\epsilon_{\kv}) a_{\kv\sigma}
 -\HA 
\end{align}
where $\HA$ describes the interaction with spatial and temporal variation of local spins, expressed by $A_\mu^\alpha$;
\begin{align}
\HA
&=
-\hbar\sum_{\kv,\qv} \left[
\sum_{\mu}
  \left(J_\mu\left(\kv+\frac{\qv}{2}\right)\cdot A_\mu^\alpha(-\qv) \right)
  a^\dagger_{\kv+\qv}\sigma_\alpha a_{\kv}  \right.\nonumber\\ 
& \left.
 +
 \frac{\hbar}{2m}\sum_{\pv}A_i^\alpha(-\qv-\pv)A_i^\alpha(\pv)
 a^\dagger_{\kv+\qv} a_{\kv}
 \right] .
\end{align}
 Here we have defined 
\begin{equation}
J_\mu(\kv)\equiv \left( \frac{\hbar}{m}\kv, 1  \right),
\end{equation}
for $\mu=x,y,z,t$.

\begin{figure}[h]
\begin{center}
{\includegraphics[width=0.4\linewidth,clip]{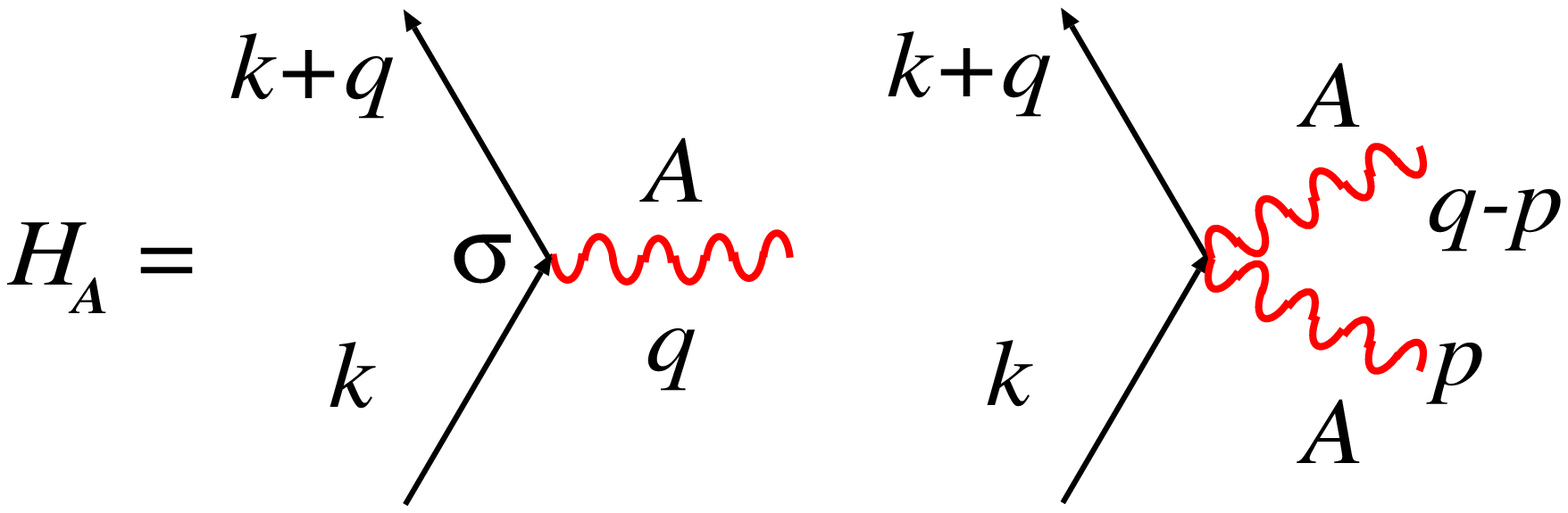}}\\ 
\vskip 2mm
{\includegraphics[width=0.4\linewidth,clip]{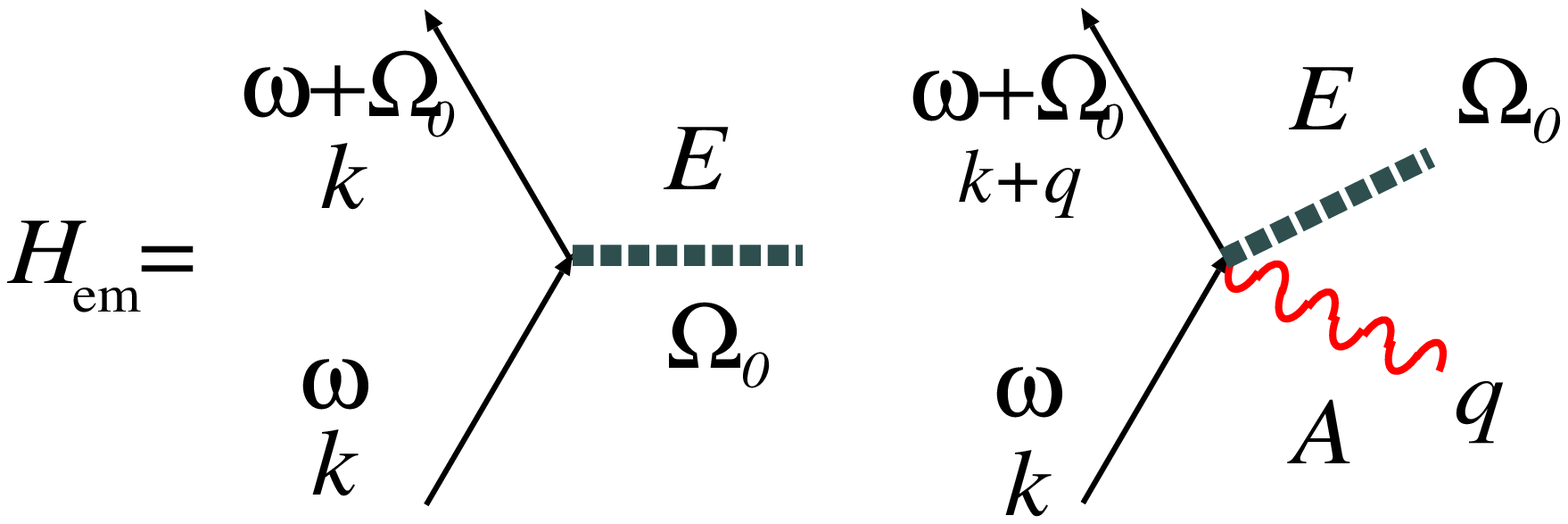}}\\  \vskip 2mm
{\includegraphics[width=0.18\linewidth,clip]{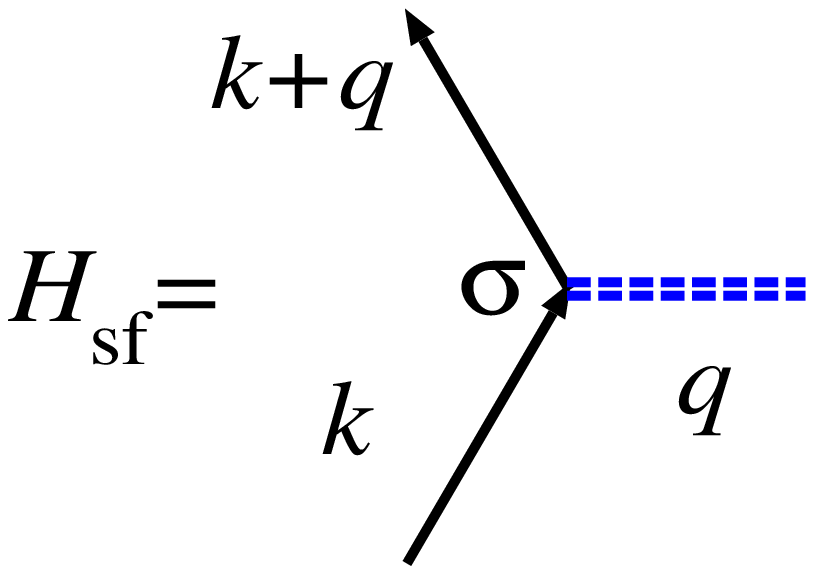}} \caption{Diagrammatic representaion of the coupling of conduction electrons to domain wall ($\HA$), to external electric field ($\Hem$), and to magnetic impurities ($\Hsf$). 
Solid lines are electron Green functions with selfenergy due to impurity scattering included, and wavy lines denote the SU(2) gauge field representing spin structure. 
}
\label{FIGinteractions}
\end{center}
\end{figure}

The total electric current, eq.(\ref{current}), is modified by the SU(2) gauge field as
\begin{align}
\Jv &=      
\frac{e}{m} \sum_{\kv} 
\lt[ (\hbar\kv - e\Avem ) a^{\dagger}_{\kv} a_{{\kv}} 
+ \hbar\sum_{\qv,\alpha}
\Av^\alpha (\qv) a^{\dagger}_{\kv+\qv} \sigma^{\alpha} a_{\kv} 
\rt]
,\label{j}
\end{align}
and the interaction with external electric field is given by
\begin{align}
\Hem &= \sum_{\kv,i}\frac{ie\hbar E_i}{m\Omz}e^{i\Omz t}
\lt[ k_i  a^{\dagger}_{{\kv}} a_{{\kv}}
+\sum_{\qv,\alpha} A^\alpha _i(\qv) a^{\dagger}_{\kv+\qv}\sigma^{\alpha}a_{\kv}
\rt] \label{hemmod}
\end{align}
up to $O(E)$. 
 To summarize the electron part, the Lagrangian is now written as
$\Le - H_{sd} = L_{\rm e}^{0}-\HA-\Hem-\Hsf$, where 
$L_{\rm e}^{0}\equiv \sum_{\kv\sigma}  a^\dagger_{\kv\sigma}
( i\hbar\partial_t -\epsilon_{\kv\sigma}) a_{\kv\sigma}-\Himp$, 
with $\epsilon_{\kv\sigma} \equiv \ekv -\sigma\spol$, 
defines free electrons under uniform magnetization and impurity potential. 
 The interactions which we treat perturbatively are 
shown in Fig. \ref{FIGinteractions}.
%

\section{Domain Wall}

 We consider a planar ({\it i.e.}, one-dimensional) domain wall as realized in a narrow wire, where the spin configuration changes only in the wire direction, which we choose as $\xw$ direction of coordinate space.
(This direction is $y$-direction of spin space in the Bloch wall case (Fig. \ref{FIGdw}). Note that spin space and coordinate space do not necessarily coincide in the symmetric system we consider.) 
 The spin part of the Lagrangian $\Ls$ (without the pinning potential and the $s$-$d$ exchange coupling) allows a static domain wall solution, 
\begin{align}
\cos\theta &=  \tanh\frac{\xw-X}{\lambda} , 
\label{dw0}
\end{align}
with $\phi=0$, and $X$ is an arbitrary constant.
Here we have introduced a length scale 
\begin{equation}
\lambda\equiv \sqrt\frac{J}{K} \label{thickness},
\end{equation}
governing the spatial scale of magnetic structure in general, 
and gives the thickness of the domain wall.

The domain wall considered above is called N\'eel wall (Fig. 1, left), where magentization is changing in the spatial $\xw$-direction, which coincides with the magnetic easy axis.
Other type of wall, called Bloch wall (Fig. 1, right), is also possible,  
if the easy plane ($zx$-plane) is perpendicular to the wire direction, $y$.
This difference of wall structure does not affect the electron transport nor the spin torque if the spin-orbit interaction is neglected.

\subsection{Collective coordinates of rigid 1D wall}
 To derive the equation of motion of a \lq rigid' domain wall, we here consider the collective coordinate description\cite{Rajaraman82}. 
 This treatment and the results are essentially the same as the one considered by  Slonczewski\cite{Slonczewski72,Hubert98} in the context of dynamics under magnetic field.

 The idea is to consider the constant $X$ in Eq.(\ref{dw0}) as a dynamical variable, $X(t)$. 
Then the angle $\phi(\xw,t)$ can be excited, too, and so another collective variable 
\begin{equation} 
 \phiz (t) \equiv \int \frac{d\xw}{2\lambda}\sin^2 \thetaz \, \phi (z,t)   \label{phizdef}
\end{equation}
needs to be treated also as dynamical\cite{TT96}, 
where $\cos\thetaz = \tanh\frac{z-X(t)}{\lambda}$ or 
$\sin\thetaz = \left[ \cosh\frac{z-X(t)}{\lambda} \right]^{-1}$. 
 This is because $X$ and $\phiz$ are canonically conjugate to each other, 
as indicated by the fact that the first term of Eq.(\ref{Ls}) takes the form $\propto \dot X \phiz$. 

In the absence of sample inhomogeneity and driving force, $X$ describes a gapless zero mode owing to the translational symmetry of the system. 
 If the pinning potential $V_0$ is present, the energy scale of $X$ will be $V_0$. 
 Similarly, the energy scale of the $\phiz$-mode is given by $\Kp$. 
 Since the energy gap of the spin-wave mode is $\sim \sqrt{KK_\perp}$, the modes described by $X$ and $\phiz$ are low energy compared to others if the following condition is satisfied; 
\begin{equation}
 \Vz \ll \sqrt{KK_\perp}, \ \ \  \Kp \ll K. 
\label{rigidity}
\end{equation}
 In this case, the low-energy wall dynamics is described by the two variables, $X$ and $\phiz$. 
 Otherwise, the pinning and/or $\Kp$ leads to a deformation of the wall, whose description requires other variables than $X$ and $\phiz$.
 The condition (\ref{rigidity}) gives a criterion that such deformations can be neglected.

 Precisely speaking, we need one more condition that there is no linear coupling of spin-wave modes to $X$ or $\phiz$. 
 In reality, when $V_0$ and $\Kp$ are finite,  there arise such linear couplings, and the wall dynamics is not closed in $X$ and $\phiz$ in a strict sense.
This is quite natural since the pinning and $\Kp$ results in a deformation of the wall whose description requires other variables than $X$ and $\phiz$.
 However, the condition (\ref{rigidity}) also assures that such linear couplings are small. 
 We assume the condition (\ref{rigidity}) in this paper.

\subsection{Domain wall Lagrangian}

From these considerations, the Lagrangian for low-energy dynamics of a rigid wall is given by using ${\bm n}_0 = (\theta_0, \phi_0)$ in the $\Ls$\cite{TT96}.
The result is $L=\Ldw+\Le$, 
where
\begin{align}
\Ldw &= \hbar NS \left( \frac{\dot{X}}{\lambda}\phiz 
   -\frac{\Kp}{2\hbar}S \sin^2\phiz \right) 
-\Vpin[\nvz]
 \nonumber\\
&
 +{\spol} \intx \, \evsz \cdot \hat\sv.
  \label{Ldw}
\end{align}
Here 
$\hat\sev\equiv c^\dagger \sigmav c$ is the electrons' spin-density operator, and $N\equiv 2\lambda A/a^3$ is the number of spins in the wall. ($A$ is the crossectional area of the wire.)

The equations of motion of the wall are now obtained simply by
taking variations with respect to $X$ and $\phiz$ and taking the expectation value of $\sv\equiv \average{\hat\sv}$.
Using
\begin{equation}
\Ws=\frac{\alphaz N\hbar S}{2}
  \left( \frac{\dot{X}^2}{\lambda^2}+\dot{\phiz}^2 \right),
\end{equation}
 in eq. (\ref{spineq}), they are obtained as 
\begin{align}
\dot\phiz+\alphaz \frac{\dot{X}}{\lambda}
 &= \frac{\lambda}{\hbar NS} (\Fe+F_{\rm pin}) , 
\label{DWeq_a}
\\ 
\dot{X}-\alphaz\lambda\dot{\phiz} 
&= \frac{\Kp\lambda}{2\hbar}S \sin 2\phiz  
   + \frac{\lambda}{\hbar NS} \tau_{{\rm e},z}.
\label{DWeq_b}
\end{align}
Here force and torque due to electrons are defined as
\begin{align}
\Fe & \equiv  - \average{\deld{H_{sd}}{X}} 
  = -{\spol}\intx \, \nabla_{\xw}\evsz \cdot\sev , 
\\
\torqueve & \equiv  -\intx \average{\deld{H_{sd}}{\Sv}}  \times \Sv
 = -{\spol}\intx  (\evsz \times\sev).
\end{align}
We note that this set of equations, (\ref{DWeq_a}) and (\ref{DWeq_b}), is essentially the same as those obtained by Berger\cite{Berger84,Berger92}.
What is new and essential in the present theory is that we have formal but exact expressions of force and torque, which we can evaluate by a systematic diagrammatic method.

Defining  each component of $\sev$ as (see Fig.2 for the definition of $\evth$ and $\evph$) 
\begin{equation}
\sev \equiv \seth\evth + \seph \evph + \sez \evs ,\label{svdef}
\end{equation}
force and torque 
are represented in terms of $\seth$ and $\seph$ as
\begin{align}
F_\mu 
 & = 
 2\spol \intx \left(-\seth A^\phi_\mu +\seph A^\theta_\mu\right)
   \label{force2}\\
\torqueve 
  &=  \spol \intx \left( \seph \evth - \seth \evph \right),
 \label{torque2}
\end{align}
where $A^\phi_\mu\equiv \evph\cdot \Av_\mu$, $A^\theta_\mu\equiv \evth\cdot \Av_\mu$.
Clearly, the component $\sez$ parallel to the local spin ${\bm S}$ does not affect the dynamics of ${\bm S}$.

\section{Calculation of Electron Spin Density}

\begin{figure}[tbh]
\label{FIGkeiro}
\begin{center}
\includegraphics[width=0.4\linewidth]{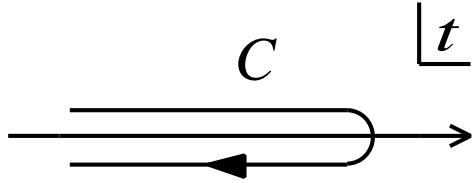}
\caption{
Keldysh contour in the complex time plane used in non-equillibrium Green functions.
}
\end{center}
\end{figure}

Our task is now to evaluate the electron spin density $\sev$ in the presence of spin structure ($\nabla \nv$) and current flow, or in the presence of spin dynamics ($\dot\nv$). 
 Since the electron state is better defined in gauge-transformed (rotated) frame, we define Green functions with respect to the rotated frame.
We use non-equillibrium (or Keldysh) Green function defined on complex time plane\cite{Haug98}, 
\begin{equation}
g_{\kv\sigma}(t,t') 
\equiv -i\average{T_C a_{\kv\sigma}(t)\adag_{\kv\sigma}(t')},
\end{equation}
where $T_C$ represents path order on complex time plane (Fig. \ref{FIGkeiro}) and $\average{\cdots}$ denotes averaging over quantum states, random impurities and thermal averaging.
This non-equillibrium Green function contains besides retarded and advanced Green functions the  lesser (greater) Green function, 
\begin{equation}
\gless_{\kv\sigma}(t,t') 
\equiv i\average{ \adag_{\kv\sigma}(t')a_{\kv\sigma}(t)},
\end{equation}
which gives directly the informaiton on the particle (hole) number and is 
is most useful in calculating physical qunatities.
%
The quantum state is defined with respect to $L_{\rm e}^{0}$, with normal impurities  taken account in the standard ladder approximation (i.e., neglecting small corrections of $O(1/(\ef \tau))$).
Thus we have free retarded and lesser Green functions as
\begin{align}
\gr_{\kv\sigma}(\omega) &= \frac{1}{\omega-\ekvs+\frac{i}{2\tau_\sigma}} 
\\
\gless_{\kv\sigma}(\omega) &= f(\omega)(\ga_{\kv\sigma}(\omega) -\gr_{\kv\sigma}(\omega) ),
\end{align}
where $f(\omega)\equiv \frac{1}{e^{\beta\omega}+1}$.

The exact Green function (detnoted by $G$), taking account of interaction, $\HA$, $\Hem$ and $\Hsf$, satisfies the same Dyson equation as the conventional time-ordered and retarded Green functions (but with time defined on a complex plane).
This Green function is still defined on complex time, $t, t'$.
To obtain physical quantities, we need to map $t$'s onto real time\cite{Haug98}.

We consider a slow local spin dynamics and assume 
that the SU(2) gauge field has only zero frequency component, $A_{\mu}^{\alpha}(\qv,\Omega) \equiv \delta_{\Omega,0} A_{\mu}^{\alpha}(\qv)$.
This is justified when $\Omega \tau \ll 1$.
We can easily evaluate the spin density in rotated frame, 
$\svtil$, defined by
\begin{equation}
\svtil(\xv,t) \equiv  \average{\adag\sigmav a} 
= -i \tr [\sigma_\alpha G^{<}(\xv,t,\xv,t)].
\end{equation}
 The spin density in the original frame is given by 
\begin{equation}
\sev(\xv,t)  =2\mv(\mv\cdot\svtil)-\svtil. \label{svdef0}
\end{equation}

\begin{figure}[tbh]
\label{FIGse}
\begin{center}
\includegraphics[width=0.3\linewidth]{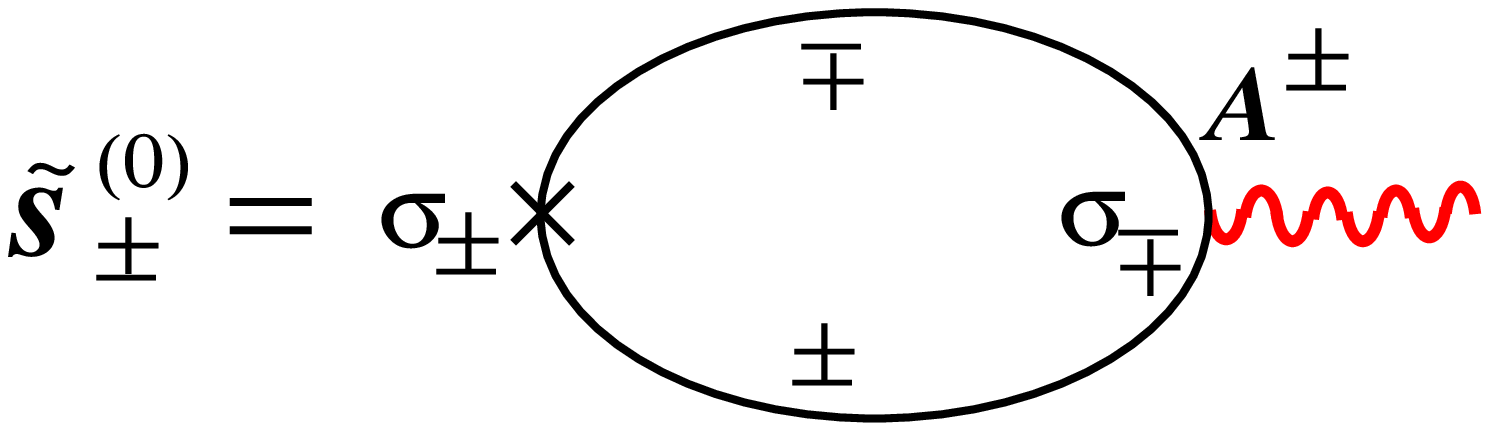}
\includegraphics[width=0.7\linewidth]{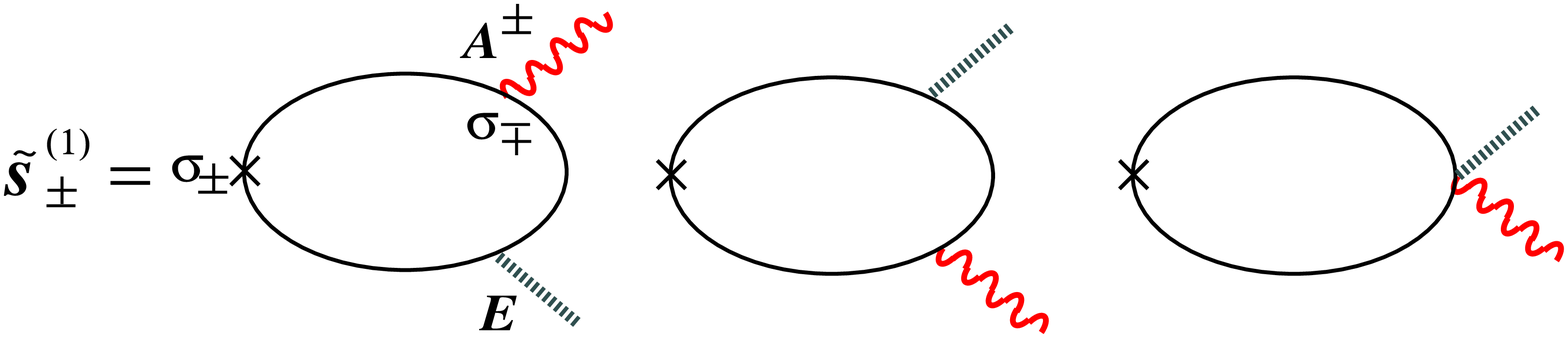}
\caption{
Contributions to the electron spin density in the rotated frame, in zeroth- and linear-oder in the applied electric field $E$ (denoted by $\tilde{\se}^{(0)}$ and $\tilde{\se}^{(1)}$, respectively).
 }
\end{center}
\end{figure}

\section{Spin-Transfer Torque on Domain Wall}

We consider a rigid one-dimensional domain wall represented by (\ref{dw0})(\ref{phizdef}). 
The correponding gauge field is given as ($q$ is along $\xw$-direction) 
\begin{align}
\Ath_0(q) &= -\frac{\pi\lambda}{2L}e^{iqX} \dot{\phiz} u_q,  \;\;\;\;
\Aph_0(q) = \frac{\pi\lambda}{2L}e^{iq X}  \frac{\dot{X}}{\lambda}u_q  \nonumber\\
\Aph_i(q) &= -\delta_{i,\xw} \frac{\pi}{2L}e^{iq X}u_q ,
\end{align}
where $u_q\equiv \frac{1}{\cosh(\frac{\pi}{2}\lambda q)}$ is a form factor of the wall and $\Ath_i = 0 $.
The electron spin polarization around the domain wall is then obtained as\cite{TKSLK07} 
\begin{align}
\seth(\xv) 
 &=
\frac{\pi\lambda}{2L\spol}\sumqv e^{-iq(\xw-X)} u_q 
\nonumber \\ 
& \hskip -3mm \times 
\left[
\se \dot{\phiz}  \chiz_1(q) 
 -\kf^3 \frac{\dot{X}}{\lambda} \chiz_2(q) 
 -\frac{j}{e\lambda} \chio_2(q) \right] , 
\label{spindw_a}
\\
\seph(\xv) 
 &=
-\frac{\pi\lambda}{2L\spol}\sumqv e^{-iq(\xw-X)} u_q 
\nonumber \\ 
& \hskip -3mm \times 
\left[
\se \frac{\dot{X}}{\lambda}  \chiz_1(q) 
 +\kf^3 \dot{\phiz} \chiz_2(q) 
 +\frac{Pj}{e\lambda} \chio_1(q) \right], 
\label{spindw_b}
\end{align}
where $\se\equiv (k_{F+}^3-k_{F-}^3)/(12\pi^2)$ is the equilibrium spin density of electrons, $\kf\equiv (k_{F+}+k_{F-})/2$, and 
$P\equiv \js/j = \deltan /n$ is the polarization of current with $n=n_++n_-$ and $\deltan=n_+-n_-$.
Dimensionless correlation functions are given by 
\begin{align}
\chiz_1(\qv) &= \frac{\hbar\spol}{\se V} \sum_{\kv,\pm} 
 {\rm P} 
\frac{f_{\kv\pm}}{\epsilon_{\kv+\qv}-\epsilon_{\kv} \pm2\spol } , 
\\
\chiz_2(\qv) &= \frac{\hbar\spol}{\kf^3 V} \sum_{\kv,\pm} 
\frac{\pi}{2}(f_{\kv+}-f_{\kv-}) 
\delta\left( \epsilon_{\kv+\qv}-\epsilon_{\kv} \pm2\spol \right) ,
\label{chizdef}
\end{align}
\begin{align}
\chio_1(\qv) &= \frac{\spol}{3\pi m \deltan V} \sum_{\kv,\pm} 
\left( \qv\cdot\left(\kv+\frac{\qv}{2}\right)\right) 
i\frac{\gr_{\kv\pm}-\ga_{\kv\pm}}
 {\epsilon_{\kv+\qv}-\epsilon_{\kv}\pm2\spol} , 
\\
\chio_2(\qv) &= \frac{\spol}{6\pi m n V} \sum_{\kv,\pm} 
\left( \pm\frac{\pi}{2} \right) 
\left( \qv\cdot\left(\kv+\frac{\qv}{2}\right)\right) 
\nonumber \\ 
& \hskip 10mm \times 
i(\gr_{\kv\pm}-\ga_{\kv\pm})
\delta( {\epsilon_{\kv+\qv}-\epsilon_{\kv} \pm2\spol} ) .
\label{chi2}
\end{align}
As seen, $\chi_1^{(i)}$ and $\chi_2^{(i)}$ ($i=0,1$) arise from the real and imaginary parts of the factor, 
$\frac{1}{\epsilon_{\kv+\qv,\mp}-\epsilon_{\kv\pm}+\frac{i}{\tau}}$, 
respectively.

Let us look into the adiabatic limit by putting $q=0$ in $\chi$'s. 
The spin density then becomes 
\begin{align}
\seth^{\rm (ad)}(\xv) 
 &=
\frac{1}{2\spol} \sin\thetaz(\xw-X)
\se \dot{\phiz} , 
\\
\seph^{\rm (ad)}(\xv) 
 &=
-\frac{1}{2\spol} \sin\thetaz(\xw-X)
\left(
\se \frac{\dot{X}}{\lambda} +\frac{Pj}{e\lambda} \right) .
\end{align}
Results (\ref{spindw_a}) and (\ref{spindw_b}) indicate that non-adiabaticity (finite $q$ contribution, $\chi_2^{(0)}$) basically exchanges roles of $\theta$ and $\phi$; non-adiabatic contribution from current induces a spin polarization in the $\theta$-direction, and drives tilt of the wall. 
This is exactly what is expected in the presence of spin relaxation as argued by Zhang and Li\cite{Zhang04}, so non-adiabaticity and spin relaxation have essentially the same effect on the spin structure.
\begin{figure}[tbh]
\begin{center}
\includegraphics[width=0.5\linewidth]{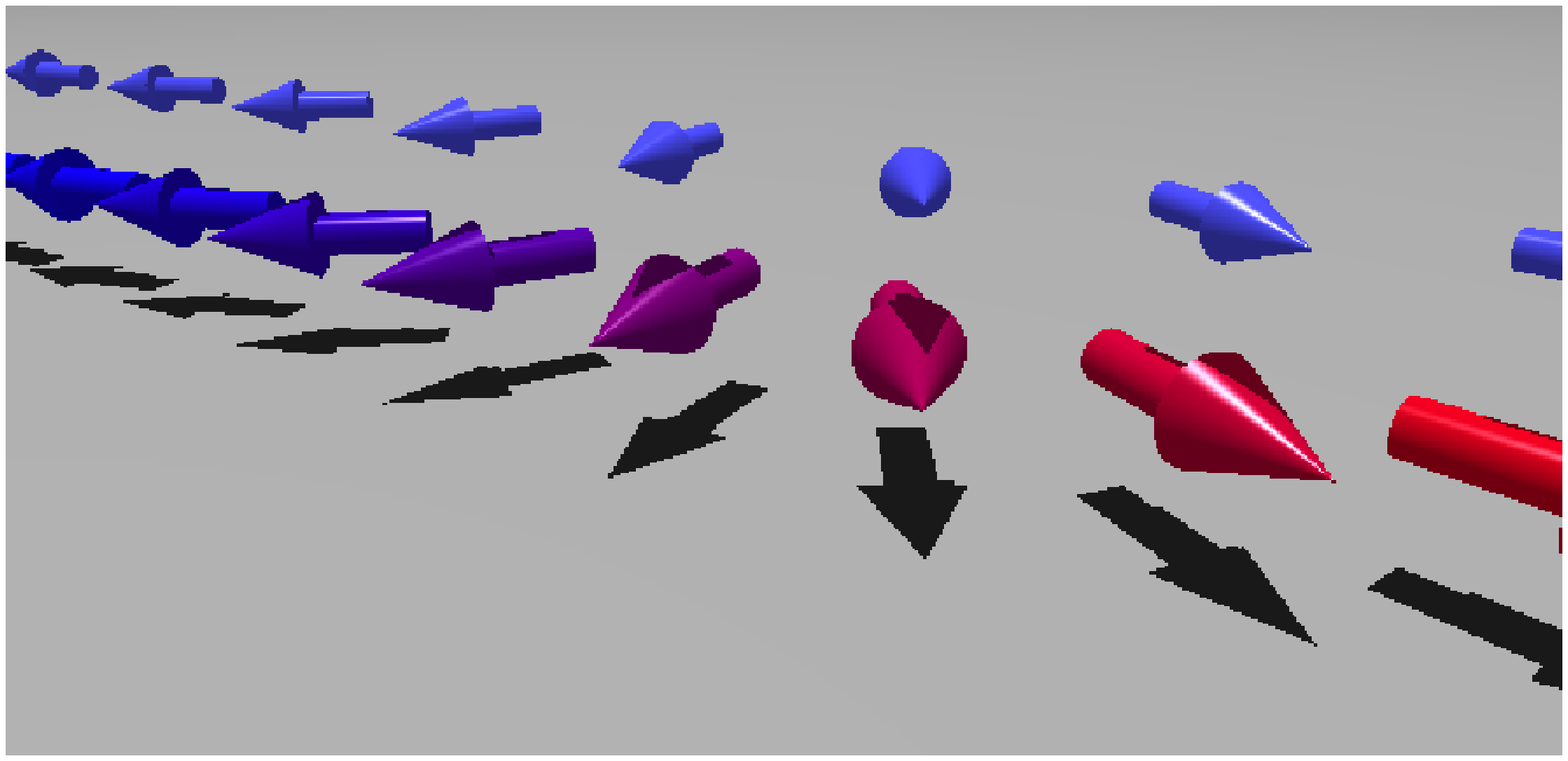}\\
\includegraphics[width=0.5\linewidth]{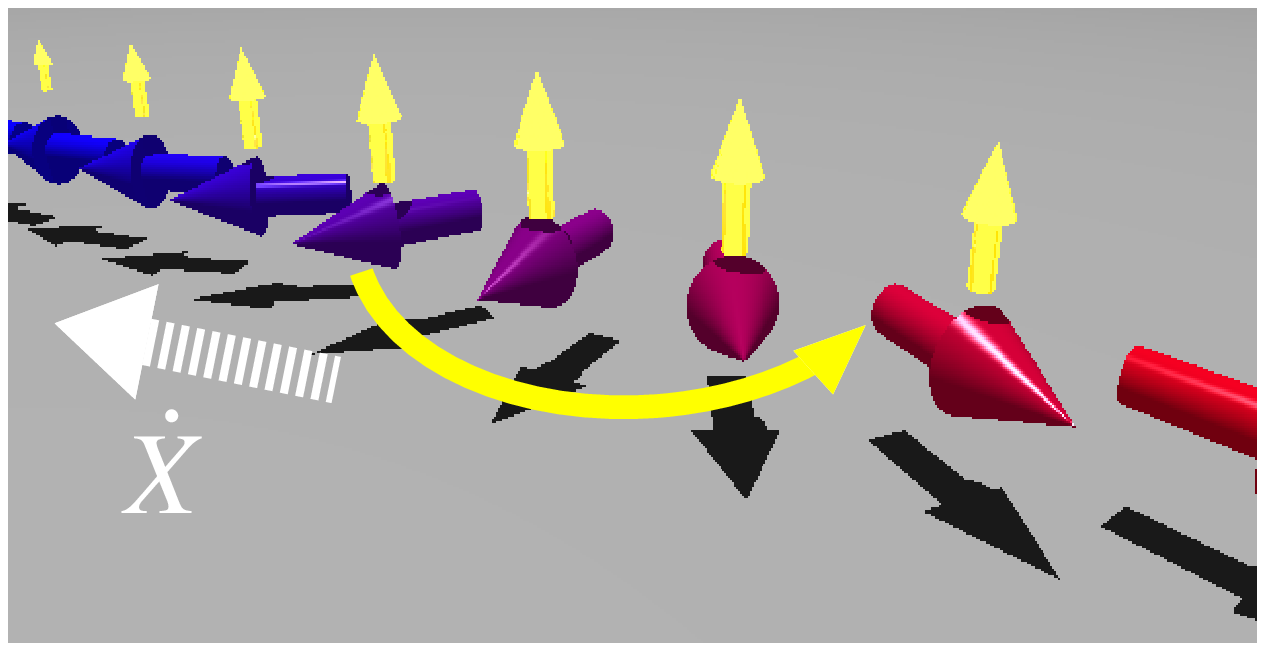}\\
\includegraphics[width=0.5\linewidth]{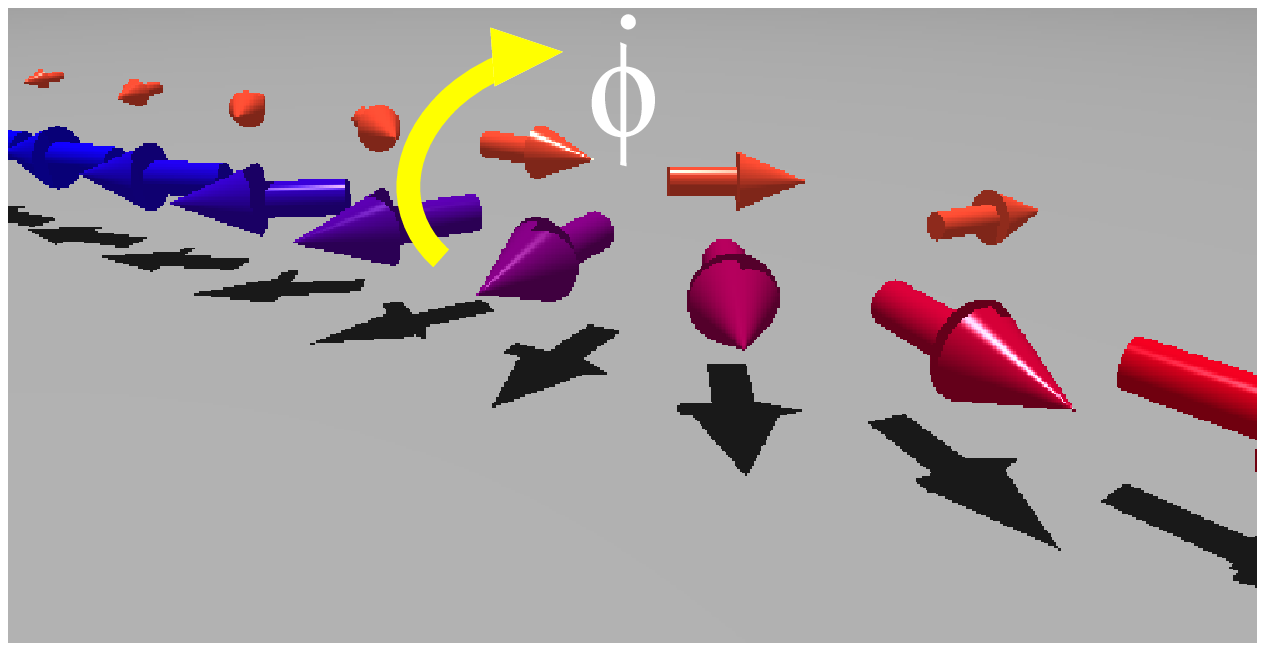}\\
\caption{ (color online)
Spin polarization of conduction electrons (denoted by small arrows) around a domain wall (large arrows), and its effect on the domain-wall dynamics.  Top: Equilibrium polarization along the local spin does not affect the dynamics. 
Middle: Under current, the spin-transfer torque is induced by the polarization component out of the wall plane (in spin space), which induces rotation of local spins within the plane, hence the translational motion, $\dot{X}$.
Bottom: $\dot{\phiz}$ is induced by the component in the $\theta$-direction in the wall plane.  Such polarization arises from non-adiabatic\cite{TK04} and spin-relaxation\cite{Zhang04} processes. }
\end{center}
\end{figure}

These features are cleary seen in the equation of motion.
The torque on a wall is obtained as
\begin{align}
\torquee 
&=
-\hf \intx\intx' \sin\thetaz(\xv) \sin\thetaz(\xv') 
\nonumber\\
& \times 
\lt[ 
\se \frac{\dot{X}}{\lambda} \chitilz_1(\xv-\xv') 
- \sum_{i} \frac{\js^i}{e\lambda}  
\chitilo_1(\xv-\xv')
\rt].\nonumber\\
&&
\end{align}
Noting $\tilde{\chi}_1^{(i)}(q)=1+O(q/\kf)^2$, we obtain  
\begin{align}
\torquee
&=
-\frac{\hbar N}{2\lambda} \lt(
\se \dot{X} - \frac{Pj}{e}  
\rt) + o((\kf\lambda)^{-2}) .\label{torquedw}
\end{align}

Roles of electron spin polarization on wall dynamics is summarized in Fig. 7.

\section{Spin Relaxation}
 The presence of spin relaxation processes in the electron system 
produces a new type of torque 
\begin{equation}
\torquev_{\rm sf} = 
 \alpha_{\rm sf} S \nv\times \dot{\nv}
+ \beta_{\rm sf} \frac{a^3 S}{2e} (\nv\times (\jsv\cdot\nabla) \nv),
\end{equation}
in the adiabatic regime, 
as first recognized in ref.\cite{Zhang04}. 
 The first term is the Gilbert damping, and the second term is 
a new type of current-induced torque which is orthogonal to the 
spin-transfer torque. 
 ($\alpha_{\rm sf}$ and $\beta_{\rm sf}$ are dimensionless 
coefficients.) 
In ref.\cite{KTS06}, the authors adopted quenched magnetic impurities 
to simulate spin relaxation processes in a microscopic Hamiltonian, 
and obtained the result as 
\begin{align}
  \alpha_{\rm sf} 
&=  \pi n_{\rm s} u_{\rm s}^2 \left[\, 
     2 \overline{S_z^2} \DOS_+ \DOS_- 
    + \overline{S_\perp^2} (\DOS_+^2 + \DOS_-^2 ) 
    \right] , 
\label{eq:alpha1}
\\
  \beta_{\rm sf} 
&=  \frac{\pi n_{\rm s} u_{\rm s}^2}{M} 
    \left[ 
    \bigl( \overline{S_\perp^2} + \overline{S_z^2} \bigr) \tilde\DOS_+ 
  + \frac{1}{P} 
    \bigl( \overline{S_\perp^2} - \overline{S_z^2} \bigr) \tilde\DOS_- 
    \right] ,
\label{eq:beta1}
\end{align}
in terms of the density of states, 
$\nu_\pm$, 
or $\tilde\nu_\pm \equiv \nu_+ \pm \nu_-$ 
(note that the notation is different from ref.\cite{KTS06}), 
the concentration ($n_{\rm s}$) and the scattering amplitudes 
($u_{\rm s} S_\perp$, $u_{\rm s} S_z$) of magnetic impurities, 
and the degree of spin polarization, $P=j_{\rm s} /j$, of the current.

\section{Force}

\begin{figure}[tbh]
  \begin{center}
  \includegraphics[width=0.6\linewidth]{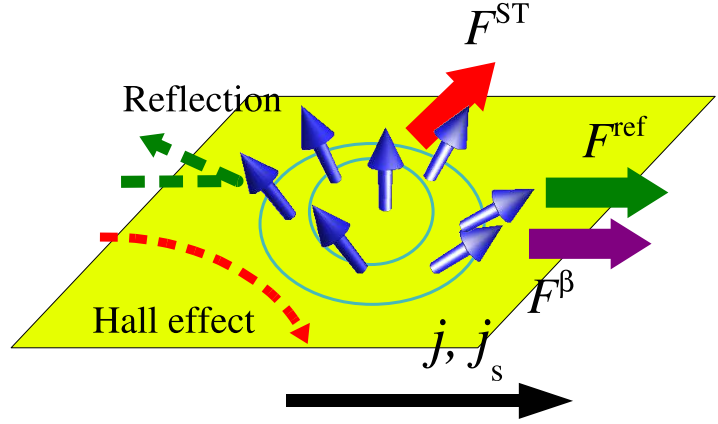}
\caption{ (color online)
Schematic illustration of three kinds of current-induced forces acting on a spin structure. 
Electron reflection pushes the structure along the current flow ($F^{\,{\rm refl}}$), and Hall effect due to spin chirality results in a force in the perpendicular direction ($F^{\,{\rm ST}}$). 
Spin relaxation results in a force along the direction of current ($F^{\beta}$).
\label{FIGvortexwall}}
  \end{center}
\end{figure}

The concept of force may be generalized to 
arbitrary spin structures based on 
Eq. (\ref{force2}).
There are several types of forces corresponding to each torque 
\cite{TKSLK07,KTSS06}. 
 In particular, we now have three kinds of current-induced forces: 
${\bm F} = {\bm F}^{\,{\rm refl}} + {\bm F}^{\,{\rm ST}} + {\bm F}^\beta$. 
 The first one 
\begin{align}
{\bm F}^{\,{\rm refl}} &= e \Ne\rhos {\bm j} , \label{forcedw}
\end{align}
is a (non-adiabatic) force due to electron reflection 
by spin structure, and is related to the resistivity\cite{TF97,GT01}
\begin{align}
\rhos &= 
 \frac{4\pi M^{2}}{e^{2}n^{2}}\frac{1}{V}
 \sum_{\kv, q, \sigma} |A_\xw^\sigma(\qv)|^{2} \, 
\delta(\epsilon_{\kv,\sigma} ) \, 
\delta(\epsilon_{\kv+q,-\sigma} )  , 
\label{rhos}
\end{align}
due to the spin structure. 
The other ones, 
${\bm F}^{\,{\rm ST}}$ and ${\bm F}^\beta$, 
are finite in the adiabatic limit, and 
are physically different from the reflection force, ${\bm F}^{\,{\rm refl}}$.
They are calculated as
\begin{align}
F^{\,{\rm ST}}_i &= 
-\frac{1}{2e} \sum_\ell {\js}_\ell \intx \, 
\nv \cdot(\partial_i\nv \times \partial_\ell \nv ) 
\\ 
F^{\,\beta}_i &= \beta_{\rm sf} \frac{1}{2e} 
\int d^3x \, 
(\jsv\!\cdot\!\nabla) \nv \cdot  \nabla_i \nv , 
\label{Fbeta}
\end{align}
up to $O((\kf\lambda)^{-2})$. 
 For ferromagnetic films, the spin-transfer force ${\bm F}^{\,{\rm ST}}$ has a topological meaning since the quantity, 
$\nvortex \equiv  \frac{1}{4\pi}  \int d^2x \, 
\nv\cdot(\partial_x\nv \times \partial_y \nv ) $, 
defined in two dimensions is a topological number 
(integer or half integer depending on boundary conditions). 
The force ${\bm F}^{\,{\rm ST}}$ is, in fact, a back reaction of the Hall effect due to spin chirality\cite{Ye99,TKawamura02}, and was derived by Thiele\cite{Thiele73} and Berger\cite{Berger86}, then based on a general relation between force 
and the torque\cite{KTSS06}, and also 
in the context of magnetic vortex\cite{SNTKO06}.
Note that the adiabatic force is included in the (adiabatic) 
spin-transfer torque, while the reflection force is not. 
The second term due to the $\beta$-term is written as  
\cite{Thiaville05,Zhang04,KTSS06}
\begin{equation}
F^{\beta}=\frac{1}{2e} \gamma'\beta_{\rm sf} \js,
\end{equation}
for a one-dimensional spin texture, 
where $\gamma' \equiv  \intx \, (\nabla_\xw \nv)^2$.
These forces are schematically illustrated in Fig. \ref{FIGvortexwall}.

 There are also forces induced by spin dynamics ($\dot {\bm S}$) 
in the absence of current, ${\bm F} = {\bm F}^{\, (0)} + {\bm F}^{\, (1)}$ :  
\begin{align}
F^{\, (0)}_i &= \frac{\se }{2} \sumx \, 
\nv \cdot (\dot\nv \times \partial_i \nv) ,
\label{F_0} \\ 
F^{\, (1)}_i &= \alpha_{\rm sf} S \sumx \, 
\dot{\nv} \cdot \partial_i \nv 
\label{F_1}
\end{align}
The first term comes from the \lq\lq spin renormalization'' torque (see the next section), and 
the second term comes from the Gilbert damping.

\section{Equation of Motion}
For a rigid one-dimensional wall, 
$\Fz=-\frac{NS}{2\lambda}\dot{\phiz}+ o((\kf\lambda)^{-2})$ and 
$F^{\,{\rm ST}} = 0$. 
(For a vortex or a vortex wall, $F^{\,{\rm ST}}$ is finite.)
Combining Eqs. (\ref{DWeq_a}), (\ref{DWeq_b}), (\ref{torquedw}) and (\ref{forcedw}), 
we finally obtain the equation of motion under current as
\begin{align}
\renom\dot\phiz+\alpha \frac{\dot{X}}{\lambda}
 &= \frac{a^3}{e\lambda} \beta j + \fpin , 
\label{DWeq3_a} \\
\renom\dot{X}-\alpha\lambda\dot{\phiz} 
&= \vc \sin 2\phiz  
+\frac{a^3}{2S}P\frac{j}{e} ,
\label{DWeq3_b}
\end{align}
up to $O((\kf\lambda)^{-2})$, 
where the damping now includes the contribution from spin relaxation (eq. (\ref{eq:alpha1})), 
$\alpha=\alpha_0+\alpha_{\rm sf}$, 
\begin{align}
\beta &\equiv \frac{\lambda}{2\hbar S} e^2n\Rw A + P\betasf
,\label{betadef}
 \end{align}
is the total force due to electric current including the effect of spin relaxation (eq. (\ref{eq:beta1})), and 
\begin{equation}
\vc\equiv
\frac{\Kp\lambda}{2\hbar}S 
\end{equation}
has the dimension of velocity. 
Pinning force represented by
\begin{align}
\fpin &= 
-\frac{2\lambda\Vz}{\hbar S \xi^2} X \theta(\xi-|X|),
\end{align}
is due to an (extrinsinc) pinning potential 
$\Vpin(X)=\frac{1}{2}\Mw \Omega^2 (X^2-\xi^2)\theta(\xi-|X|)$, which we approximate as harmonic with range $\xi$, pinning frequency $\Omega$ ($\Mw\equiv N/(\Kp\lambda^2)$ is the wall mass and $\Vz\equiv \frac{\Mw}{2N}\Omega^2\xi^2$).
The factor of $\renom\equiv \lt(S+\frac{\se}{2}\rt)/S$, arising from $\Fz$, indicates that electron spin polarization $\se/2$ contributes to the magnetization of the wall (if close to the adiabatic limit). This natural result indicates consistency of our calculation.
(In the equaitons of motion, we have neglected small non-adiabatic corrections of $o((\kf\lambda)^{-2})$.
Correctly speaking, the reflection force is also the same order of $o((\kf\lambda)^{-2})$\cite{Cabrera74,TF97,GT01}, but we retain this term since physical meaning is clear.)

Most important part of our work is the derivation of this equation of motion. 
We see that it is essentially the same as the one argued by Berger\cite{Berger84,Berger92} (with additional terms introduced by Zhang and Li\cite{Zhang04}, 
and Thiaville et al.\cite{Thiaville05}), indicating his deep physical insights.
The phenomenological arguments were quite useful in discussing the adiabatic limit, where angular momentum conservation (adiabatic spin-transfer torque) governs the dynamics.
Once non-adiabaticity and spin relaxation come in, fully quantum mechanical calculation as we did is required.
Of particular future interest would be the quantitative first-principle estimations of torques and forces including material parameters (such as spin-orbit interactions) and geometry (pinning) based on the present formulas (\ref{force2}) and (\ref{torque2}).

\section{Solution}

Let us look into the solution of the equations of motion, 
which are given in terms of dimensionless parameters by  
\begin{align}
{\partial_\ttil} \left( {\Xtil} -\alpha {\phiz} \right)
 &= \sin 2\phiz + \Ptil \, \jtil , 
\label{DWeq4_a} 
\\ 
{\partial_\ttil} \left( {\phiz} + \alpha{\Xtil} \right)  
&= 
  -\Vztil \Xtil\theta(\xi/\lambda-|\Xtil|) +\beta \, \jtil, 
\label{DWeq4_b}
\end{align}
where
$\ttil\equiv t \vc/\lambda$, $\Xtil\equiv X/\lambda$, 
$\Omegatil\equiv \Omega \lambda/\vc 
 = \frac{2\sqrt{2}}{S}\frac{\lambda}{\xi}\sqrt{\frac{\Vz}{\Kp}}$, 
$\Ptil\equiv \frac{P}{2S}$, 
$\jtil\equiv \frac{a^3}{e\vc}j$ and 
$\Vztil\equiv \frac{1}{2}\Omegatil^2
 = \frac{4}{S^2} 
 \frac{\Vz}{\Kp}\left(\frac{\lambda}{\xi}\right)^2$,
 and we approximated as $\renom=1$.
(For details of the solutions, see Ref. \cite{TTKSNF06})

\subsection{Threshold current}
Behavior of threshold current depends on the extrinsic pinning.
\paragraph{(I) Weak pinning regime}

Under small current, $\jtil\lesssim 1$, $\phiz$ remains small and the wall dynamics is well described by $X$ only. 
This is defined as regime I.
Linearizing the sine-term in Eq.(\ref{DWeq_b}) as $\sin 2\phi \simeq 2\phi$, we eliminate $\phi$ to obtain \cite{SMYT04,TSIK05}
\begin{equation}
(1+\alpha^2)\partial_{\ttil}^2 {\Xtil}
 +\frac{1}{\tautil}\partial_{\ttil}{\Xtil}
+\Omegatil^2 \Xtil = \tilde{F}_{\rm j},
\label{DWeq2}
\end{equation}
where 
$1/\tautil=2\alpha\left(1+\frac{1}{2}\Vztil\right)$, 
and 
$\tilde{F}_{\rm j}\equiv 2\beta \jtil$ 
is a dimensionless force due to current.
We consider the case of steady current and weak damping; 
$2\Omegatil\tautil > 1$. 
A solution satisfying the initial condition, $\Xtil(0)=0$ 
and $\partial_{\ttil}\Xtil(0)=\Ptil \jtil$, is obtained as 
\begin{align}
\Xtil(t) &=  \frac{2\beta \jtil}{\Omegatil^2}
\left(1-e^{-\frac{\ttil}{2\tautil}}
 \left(\cos\Omegaptil \ttil 
 +\frac{1}{2\Omegaptil \tautil} \sin \Omegaptil \ttil \right) \right) 
\nonumber \\ 
& \hskip 30mm 
+\frac{\Ptil \jtil}{\Omegaptil} \, 
e^{-\frac{\ttil}{2\tautil}} \sin \Omegaptil \ttil,
\end{align}
where 
$\Omegaptil\equiv \sqrt{\Omegatil^2-\frac{1}{4\tautil^2}}$.
The threshold (depinning) current is determined by $|\Xtil(\jtil)|_{\rm max}=\xi$, which is given by 
\begin{align}
{\jc}^{\rm Ia)} & \sim 
\frac{2\sqrt{2}S} {P}\frac{e\lambda}{\hbar a^3} \sqrt{\Kp\Vz} 
  & (\beta_{\rm c}\lesssim \frac{\Ptil}{2}\Omegatil)
\nonumber\\
{\jc}^{\rm Ib)} & = \frac{eS}{\hbar a^3} \frac{\lambda^2}{\xi}
 \frac{\Vz}{\beta}    & (\beta_{\rm c}\gtrsim \frac{\Ptil}{2}\Omegatil).\label{jcIb}
\end{align}
By balancing the pinning force and the force due to magnetic field, $\Vz$ is expressed by the depinning field $\Bc$ as
\begin{equation}
\Vz= \frac{S}{2}g\muB \Bc \frac{\xi}{\lambda}. \label{VzB}
\end{equation}

\paragraph{(II) Intermediate regime}
This regime, $\jtil\gtrsim O(1)$, could be important for application since the threshold is not sensitive to sample irregularities.
Depinning in this regime is described by $\phiz$ \cite{TK04}.
The reason is that the effective mass of \lq\lq $\phiz$-particle", given by $1/\Vz$\cite{TT96}(see Eq. (\ref{phieq})), becomes lighter than the corresponding mass of \lq\lq $X$-particle" given by $1/\Kp$, and so \lq\lq $\phiz$-particle" is a better variable to describe dynamics for strong pinning. 
By eliminating $X$ from Eqs. (\ref{DWeq4_a}) and (\ref{DWeq4_b}), we obtain
\begin{align}
& (1+\alpha^2) \partial_{\ttil}^2 \phiz
+ \alpha\partial_{\ttil}{\phiz}
\left(2\cos2\phiz +\Vztil\right) 
\nonumber \\ 
& \hskip 30mm 
+\Vztil\sin2\phiz + \jtil\Vztil\Ptil=0.
\label{phieq}
\end{align}
Thus $\beta$ does not affect the dynamics of $\phiz$.
(Correctly speaking, this feature is specific to the harmonic pinning potential, and anharmonicity introduces the dependence on $\beta$.) 
In fact, the $\beta$-term can be eliminated from the equations of motion if one rewrite Eq. (\ref{DWeq4_b}) in terms of 
$X'\equiv X-\frac{2\beta}{\Omegatil^2}\jtil$ 
(i.e., it just shifts the stable point of $X$).
Even in the case with anharmonicity, we have numerically checked that $\beta$ does not lead to important modification in this regime. 

From Eq. (\ref{phieq}), we see that the energy barrier for $\phiz$ vanishes when
$\jctil\sim \Ptil^{-1}$, irrespective of the pinning strength.
Once $\phiz$ escapes from the local minimum, its velocity is given from Eq. (\ref{phieq}) as
$
\partial_{\ttil}\phiz \simeq \frac{\jtil\Ptil}{\alpha}.
$
 This corresponds, according to Eq. (\ref{DWeq4_b}), 
to a maximum displacement,  
$\Xtil_{\rm max} \simeq \frac{\jtil\Ptil}{\alpha\Vztil}$, 
of the wall. 
Unless the pinning is extremely strong, i.e., if $\alpha\Vztil \lesssim 1$, 
$|\Xtil_{\rm max}|$ exceeds $\xi/\lambda$, i.e., depinning of $X$ 
occurs as soon as $\phiz$ is depinned.
Thus the threshold is roughly given by $\jctil\sim \Ptil^{-1}$,
and is actually found numerically to be 
\begin{equation}
\jc \sim 0.7\times \frac{e}{\hbar} \frac{\lambda S^2}{P a^3} \Kp.\label{jcII}
\end{equation}

\begin{figure}[tbp]
\begin{center}
\includegraphics[width=0.6\linewidth]{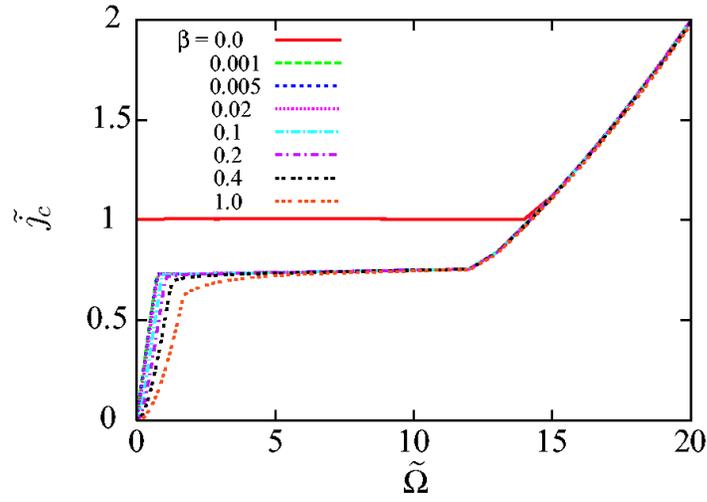}
\caption{(Color online) Threshold current $\jctil$ plotted as function of pinning frequency $\Omegatil\equiv\sqrt{2\Vztil}$ for $\alpha=0.01$, $\Ptil=1$, and several values of $\beta$. 
\label{FIGjc} }
\end{center}
\end{figure}

This story was presented in ref. \cite{TK04}, but the estimate of threshold current there was $\jctil=1$.
The reason for this difference comes from $\beta$. 
In the analysis of ref. \cite{TK04}, where $\beta=0$ was assumed, 
even if $X$ escapes from the pinning potential for current 
$\jtil > 0.7\Ptil^{-1}$, 
the terminal velocity vanishes if $\jtil <\Ptil^{-1}$ 
owing to the intrinsic pinning effect (i.e., $\phiz$ reaches a steady value and $\dot{X}$ becomes zero).
On the other hand, if $\beta\neq 0$, steady motion of $X$ is possible as soon as $X$ escapes from the pinning.
This is the reason why the threshold curreent is different for $\beta=0$ and $\beta\neq0$ in the intermediate regime II
(Fig. \ref{FIGjc}).

\paragraph{(III) Strong pinning regime}
The above result $\Xtil_{\rm max} \simeq \frac{\jtil\Ptil}{\alpha\Vztil}$ indicates that for extremely strong pinning, 
$\Vztil \gtrsim \alpha^{-1}$, the wall is not always depinned even after $\phiz$ escapes from a potential minimum due to $K_\perp$. 
 The depinning occurs at
\begin{equation}
\jctil\sim \frac{\alpha\Vztil}{\Ptil}\frac{\xi}{\lambda},
\label{jcIII}
\end{equation}
as pointed out in ref. \cite{TK04}.

\begin{table}[tb]
\caption{Summary of threshold current for different strength of extrinsic pinning. 
\label{table}}
\begin{center}
\begin{tabular}{|c|c|c|} \hline
  Pinning &     & Threshold   \\ \hline
I-a: Weak & $\Omegatil \lesssim O(1)$, $\beta\lesssim O(\Omegatil)$ & $\jc\propto \sqrt{\Kp\Vz}$ \\ \hline
I-b: Weak & $\Omegatil \lesssim O(1)$, $\beta\gtrsim O(\Omegatil)$ & $\jc\propto {\Vz}/\beta$ \\ \hline
II: Intermediate & $O(1) \lesssim \Omegatil \lesssim O(\alpha^{-1})$ & $\jc\propto \Kp$  \\ \hline
III: Strong &  $O(\alpha^{-1}) \lesssim \Omegatil$  & $\jc\propto {\Vz}/\alpha$ \\ \hline
\end{tabular}
\end{center}
\end{table}
Results for the threshold current are summarized in table \ref{table}.
It is interesting that such a simple set of equation of motion results in so rich behaviors.

\subsection{Wall speed}
After depinning, the wall dynamics is describged by the equations of motion, (\ref{DWeq4_a}) and (\ref{DWeq4_b}), with $\Vztil=0$. 
 The solution can be obtained analytically (see Eq.(31) in Ref.\cite{TTKSNF06}). 
We see that the wall dynamics is quite different for $\jtil \geq \jatil$ and $\jtil \leq \jatil$, where 
\begin{equation}
\jatil \equiv  \left| \Ptil - \frac{\beta}{\alpha} \right|^{-1} .
\end{equation}
Above $\jatil$, the wall velocity $\dot X$ has an oscillating component, while the wall reaches a steady motion below $\jatil$.
The time-averaged velocity is given by 
\begin{equation}
\average{\dot{X}} = \frac{\beta}{\alpha} \jtil
 +\frac{\sgn[(\Ptil-\frac{\beta}{\alpha})\jtil]}{1+\alpha^2}
\sqrt{\left[\left(\Ptil-\frac{\beta}{\alpha}\right)\jtil\right]^2-1} 
\end{equation}
for $\jtil \geq \jatil$, and 
\begin{equation}
\average{\dot{X}} = \frac{\beta}{\alpha} \jtil 
\end{equation}
for $\jtil \leq \jatil$.

\begin{figure}[tbp]
\begin{center}
\includegraphics[width=0.6\linewidth]{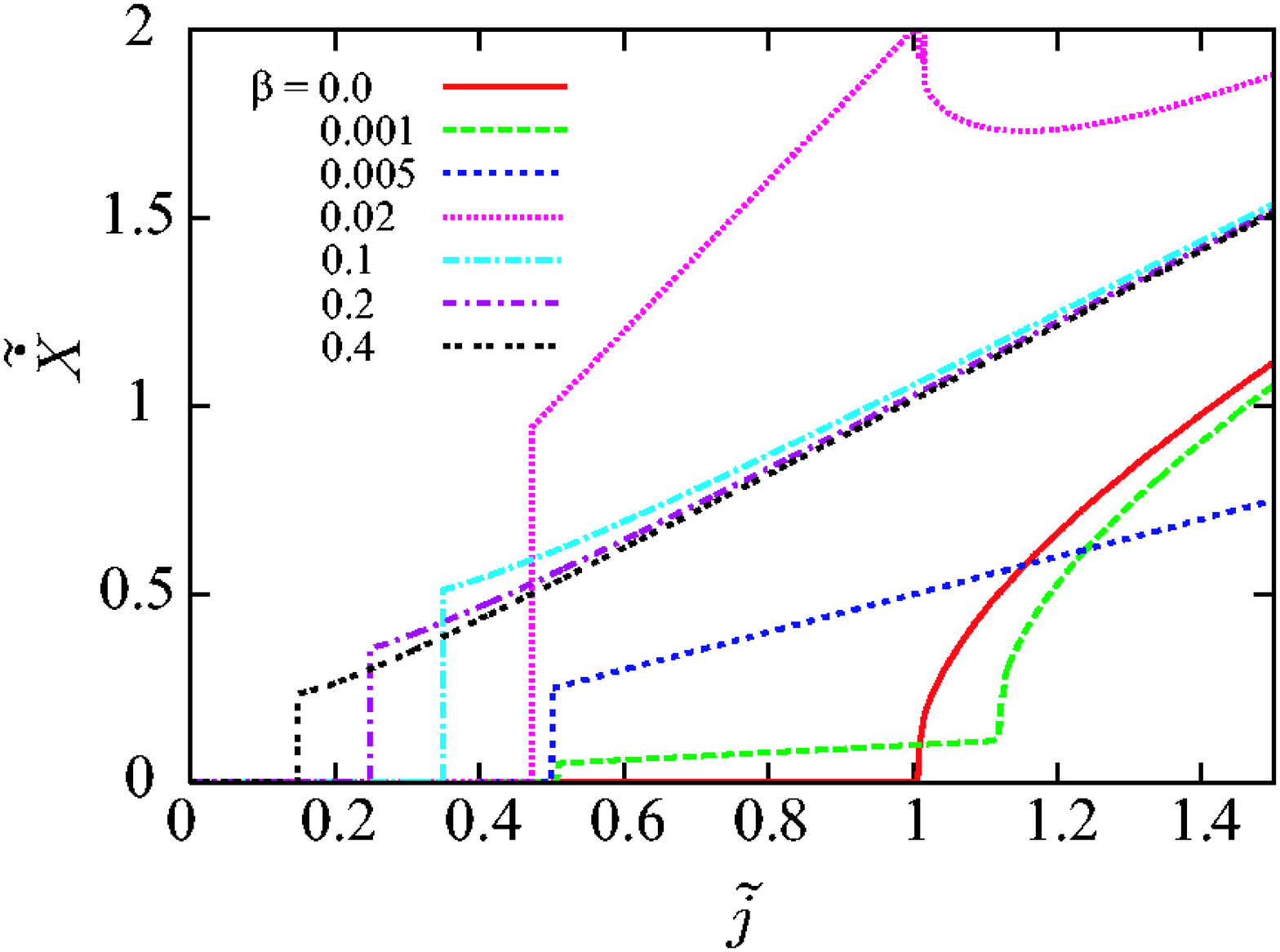}
\caption{(Color online) Wall velocity as function of current density for $\Omegatil=0.5$ and $\alpha=0.01$.
A jump is seen at $\jtil=\jctil$.
Crossover from (I-a) $\jctil\sim \Omegatil/\beta$ to 
(I-b) $\jctil \sim \Omegatil$ is seen at 
$\beta\simeq \frac{\Omegatil}{4}\sim 0.1$. 
\label{FIGvj} }
\end{center}
\end{figure}

\section{Numerical Simulation}
The above results are for a one-dimensional (1D) rigid wall, which would not be the case in real experiments (in, e.g., a thick wire of width $\Lp\gtrsim \lambda$).
Nevertheless, rigid 1D wall description seems quite good, if we compare with numerical simulation carried out on realistic sample geometries\cite{Thiaville04,Thiaville05}.
The simulations are based on Landau-Lifshitz-Gilbert equation 
with spin-transfer torque and the $\beta$-term included, 
without assuming rigid nor 1D. 
It was found there firstly that the wall speed is correlated with the appearance of hard-axis component of spin (i.e., structure like vortex core) inside the wall\cite{Thiaville04}.
In fact, during slow motion of the wall, a vortex core is nucleated, and then the wall is accelerated.  In due course, the core is annihilated, emitting spin waves, and then the wall slows down and sometimes stops. 
This oscillation of wall speed synchronized with creation and annihilation of vortex core are the same as predicted in rigid 1D case (Eqs. (\ref{DWeq3_a}) and (\ref{DWeq3_b}), where vortex core is simulated by $\phiz$ and periodic modulation of wall speed is represented by $\sin2\phiz(t)$ term in the velocity.
Thus, somewhat surprisingly, deformation and details of spin structure does not affect the dynamics in an essential way, resulting in quite a similar wall speed as a function of current (Fig. \ref{FIGvj} and Fig. 2 of Ref. \cite{Thiaville04} and Ref. \cite{Thiaville05}). 
This is because of adiabatic wall, where the spin-transfer torque tends to flow spin structures at the same velocity 
 $\frac{a^3}{2eS}Pj$ irrespective of spin structure.

Effects of deformation would be significant in the presence of extrinsic pinning, and will affect the threshold current.

\section{Recent Experiments}

\subsection{Metallic systems}

So far experimental results on metallic samples all show threshold currents of order of $10^{12}$[A/m$^2$].
If we use $\Kp/\kB\sim O(1)$ [K] estimated experimentally,\cite{Yamaguchi04,Yamaguchi05,Yamaguchi06} 
the observed threshold is orders of magnitude ($10^{-2}-10^{-1}$ times) smaller than the intrinsic threshold, $\jci$.
For instance, a sample of Yamaguchi\cite{Yamaguchi04,Yamaguchi05,Yamaguchi06} showed 
$\jc=1\times 10^{12}$ [A/m$^2$]. 
 The anisotropy energy is estimated to be $\Kp/\kB=2.4$[K], and using
$S\sim \frac{1}{2}$, $a\simeq 2.2$\AA\ and $P\sim O(1)$, we obtain 
$\jci=5.8\times 10^{13}$ [A/m$^2$], i.e., 
$\jc/\jci\sim 0.02$. 
 The observed low threshold currents in metals thus should be regarded as due to an extrinsic pinning in regime I-a) or I-b). 
 Actually, direct evidence excluding intrinsic pinning in permalloy wires so far was given by Yamaguchi et al.\cite{Yamaguchi06_jjap}.
 They prepared permalloy wires with different geometry, and relaized different perpendicular anisotropy energies  
$\frac{S^2}{a^3}\Kp\simeq (0.1-7.6)\times 10^{5}$ J/m$^3$, 
which corresponds to $\Kp/\kB\simeq 0.03-2.4$K
(per 1 spin).
The intrinsic pinning, Eq. (\ref{jcII}),  predicts then
threshold current of $5\times 10^{11}-4\times 10^{13}$ A/m$^2$.
In contrast to this much difference in the predicted values of the intrinsic threshold current, 
experimental values of threshold for these samples do not vary so much,
$(3-8)\times 10^{11}$A/m$^2$, and are smaller than the predicted intrinsic threshold by factor of 2 to 100.
Besides, data by Yamaguchi et al. indicate that these experimental values do not scale with $\Kp$, although there is a weak dependence on $\Kp$. 
Therefore the observed threshold would be of some extrinsic origin.

Let us 
first try to explain experimental result\cite{Yamaguchi05} assuming regime I-a).
Assuming $\xi\sim \lambda$, the pinning potential is estimated from the measured depinning field 
$\Bc=0.01-0.1$[T] as 
$\Vz= 0.34\times (10^{-2} \sim 10^{-1})$ [K]
$=4.7\times (10^{-26} \sim 10^{-25})$ [J], i.e.,
$\frac{\Vz}{\Kp} = 1.4\times (10^{-3} \sim 10^{-2})$, and so 
${\jc}^{\rm Ia)} = (0.21\sim0.67)\times \jci$. 
This value is still too big to explain the experimental value. 
Velocity jump is estimated as\cite{TTKSNF06} 
$\Delta v^{\rm Ia)} =\frac{\beta}{\alpha}\times 839$[m/s], so extremely small $\beta$ ($\frac{\beta}{\alpha}\sim 4\times 10^{-3}$) is required to explain the experimental value of $\Delta v\sim 3$[m/s]\cite{Yamaguchi04}.
If we assume regime I-b), the threshold is
${\jc}^{\rm Ib)} = \frac{1}{|\beta|} \times 
 2.8 \times (10^{-3} \sim 10^{-2}) \times \jci$.
Experimental value could be reproduced if $\beta= 0.1 \sim 1$.
But such large value of $\beta$ cannot be explained within the current understanding that $\beta$ arises from either non-adiabaticity\cite{TK04,Thiaville05} or spin relaxation\cite{Zhang04}.
Instead, $\Delta v$ cannot be explained by use of the above $\Vz$ assuming I-b), as it predicts too large value of 
$\Delta v^{\rm Ib)}=10^3$[m/s].
Thus, honestly, none of the above predictions based on rigid 1D wall neglecting temperature rise due to heating are successful in explaining experimental result of metals quantitatively.

It might be crucial to treat the wall as a non-rigid, non-planar object, in particular considering the sample width larger than 100nm. 
In fact, direct observation of the spin structure indicates that the wall is quite deformed upon motion\cite{Klaui05,Togawa06a,Togawa06b,Biehler07}. 
It was shown\cite{Klaui05} that the initial state 
is not a planar wall but more like a vortex
 (called a vortex wall), 
which is the case in film or wide wires, and vortex wall moves by applying a current pulse of $2.2\times10^{12}$A/m$^2$, and that the wall is deformed to be a transverse one after some pulses.
What was quite interesting there is that while vortex wall moves easier, the transverse wall does not move at the same current density.
Thus the experimentally observed wall motion in wide metallic wires would be that of vortex wall, and so simple theory assuming 1D rigid wall may not directly apply.
However, as we discussed, non rigid and non planar nature does not seem essential if we compare the results to those of numerical simulation\cite{Thiaville05}.
Threshold current of vortex wall obtained in simulation is still too large compared with experiments\cite{Nakatani07}.

There are some possibilities to resolve the disagreement. Most probable one would be the heating effect by current. 
Estimate of $\Vz$ by use of experimental $B_c$ could be an over estimate 
if effective barrier height $\Vz$ is greatly reduced by heating under current, while such heating does not occur under static magnetic field.
Let us estimate the pinning potential which gives the experimental 
value of $\jc$.
Assuming regime I-a), experimental value of $\jc/\jci=0.02$ is reproduced if
$\mu\equiv \frac{\Vz}{\Kp}=1.3\times 10^{-5}$, which corresponds to 
$\Vz=3\times 10^{-5}$[K]$=4.5\times 10^{-5}$[T].
This is two orders of magnitude smaller than the value extracted from $B_{\rm c}$.
For I-b), we have $\mu=\beta\times 10^{-2}$.
From the experiment, 
$\Delta v /(\frac{a^3}{e}\jc) = 3$[m/s]$/67$[m/s]$=0.05$. 
This value is equal, for regime I, to $\frac{\beta}{\alpha}$, so
$\beta=5\times 10^{-4}$ if $\alpha=0.01$.
So in case I-b), $\mu=5\times 10^{-6}$.
Thus, assuming either regime I-a) or I-b), the experimental results could be explained by an extremely weak pinning potential,
$\frac{\Vz}{\Kp}=10^{-6}\sim 10^{-5}$.

Heating effect in metallic samples has indeed been found to be crucially important\cite{Yamaguchi05}.
Use of short pulsed current of ns order could be useful in avoiding heating.
Sub ns pulse was reported to be quite efficient in driving the wall at low current density of $\sim 10^{10}$A/m$^2$\cite{Lim04}.
This could be due to the fact that damping does not affect much for such short timescale.

Quite recently, Dagras et al. measured the temperature dependence of the threshold current and found that it decreases at low temperatures, for instance, from 
$2.4\times 10^{12} \Ams$  at $T\sim 170$K to 
$1.9\times 10^{12} \Ams$  at $T\sim 100$K to 
\cite{Laufenberg06}.
Dissipation of spin-transfer torque by spin waves was suggested as a possible explanation, but theoretical study is yet to be done.

\subsection{Thin wall}

Quite an interesting result was obtained recently by Feigenson et al.\cite{Feigenson07} in
SrRuO$_3$, an itinerant ferromagnet with perovskite structure.
The current density needed to drive wall was 
$5.3\times 10^9\Ams$ at $T=140$K and  $5.8\times 10^{10}\Ams$ at $T=40$K.
Small threshold current at 140K would be due to reduction of magnetization close to $\Tc=150$K. 
The threshold current is about 2 orders of magnitude smaller than in other metals. 
This high efficiency would be due to a very narrow domain wall, $\lambda\sim 3$nm, as a result of very strong uniaxial anisotropy energy ($K$) corresponding to a field of 10T.
They defined a parameter determining the efficiency
as a ratio of depinning field and depinning current density, 
$\Lambda\equiv \Bc/\jc$.
Their results were $\Lambda=10^{-12}$ Tm$^2$/A.
They compared this value with threshold current of extrinsic pinning\cite{TK04}, given by Eqs. (\ref{jcIb}) and (\ref{VzB}).
Using $\frac{\hbar a^3}{e\mub\lambda}\sim 0.5\times 10^{-11}$ [Tm$^2$/A] and $S\sim 3/2$, we see that $\Lambda=10^{-12}$ Tm$^2$/A is realized if $\beta\sim 0.5$.
This value would be too large if interpreted as due to spin relaxation.
Using the measured resistivity of domain wall, the non-adiabatic force contribution to $\beta$ was estimated and the result of $\jc$ was of similar order as observed ones but with discrepancy of factor of  around 6 at low temperature 
(Fig. 4(a) of Ref. \cite{Feigenson07}). 
This discrepancy seems not so bad considering crude rigid and planar approximation of the wall.
There is another extrinsic pinning threshold, 
$j_{\rm c}^{\rm Ia)}$. 
If we use this expression,
$\Lambda=10^{-12}$ Tm$^2$/A is obtained if 
$\Kp\sim \mub\Bc=4\times(10^{-3}-10^{-2})$K (per site).

\subsection{Magnetic Semiconductor}

Beautiful experiments were carried out at low current in ferromagnetic semiconductors by Yamanouchi et al\cite{Yamanouchi04,Yamanouchi06}.
They fabricated a well structure of 20$\mu$m width made of GaMnAs with different thickness, 
 which determines the ferromagnetic coupling and transition temperature, 
and trapped a domain wall. 
The wall position was measured optically after applying a current pulse, and the average velocity was estimated.
The current necessary was $\sim 4\times 10^{9}$A/m$^2$, which is 2-3 orders of magnitude smaller than in metallic systems.
This is due to the small average magnetization, $S\sim 0.01$, carried by dilute Mn ions, and small hard-axis anisotropy $\Kp$ \cite{Yamanouchi06}.
The obtained velocity was rather high, $\sim 22$m/s at $j=1.2\times 10^{10}$A/m$^2$.
This velocity is consistent with the adiabatic spin-transfer mechanism, and the threshold appears to be consistent with intrinsic pinning mechanism\cite{TK04} with anisotropy energy obtained from band calculation. 
 However, there are some puzzles. 
 First, the theory of intrinsic pinning\cite{TK04} and adiabatic spin transfer does not take account of strong spin-orbit interaction in semiconductors. So the agreement with these thoeries might be a coincidence. 

The second puzzle is the validity of using purely adiabatic theory. In fact, quite a large momentum transfer (force) is expected from the wall resistance, $\Rw=1\Omega$\cite{Chiba06}, corresponding to $\beta\gg 1$ in terms of $\beta$\cite{Chiba06}.

The other puzzle, which was solved just recently, is the temperature dependence of wall velocity. 
The observed velocity scaled as 
$\ln v \simeq -(\Tc-T)^2 j^{-1/2}$, similar to the creep behavior under magnetic field\cite{Lemerle98}, but this fractional power of $j$ has not been explained in the current-driven case. 
A simple theory of thermal activation assuming rigid wall under the spin-transfer torque predicts different behavior, $\ln v \simeq j/T$\cite{TVF05}, and thus creep motion would be essential in the experiment by Yamanouchi et al.
Sucessful explanation of creep behavior was just recently done by Yamanouchi et al.\cite{Yamanouchi07}, by taking account of 
growth of $\phi$ at the pinning center.

Nguyen et al. studied theoretically the domain wall speed in magnetic semiconductors based on the 4-band Kohn-Luttinger Hamiltonian\cite{Nguyen07}.
It was shown there that the wall speed can be enhanced by the spin-orbit interaction by a factor of $10^{3}-10^{4}$ due to the increase of mistracking, hence reflection, of conduction electrons.
This could be useful for efficient magnetization switching.

\subsection{Excitation of wall} 
Time-resolved study of excitation of wall provides rich information on the wall character and driving mechanism.

Under AC current, domain wall shows another aspect not seen in DC case.
AC current can drive domain walls quite effectively at low current if the frequency is tuned close to the resonance with the pinning frequency. 
This resonance was realized in recent experiment by Saitoh {\it et al.}\cite{SMYT04}.
 They applied a small AC current (of amplitude of $10^{10}$A/m$^2$) in a wire with a domain wall in a weak pinning potential controlled by magnetic field.
Although the current is well below the threshold, the wall can shift slightly as we see below (for about a distance of $\mu$m, but this would be an overestimate).
Under small current, $\phiz$ remains small, and the equation of motion reduces to that of a \lq\lq particle";
\begin{equation}
\Mw\ddot{X}+\frac{\Mw}{\tau}\dot{X}+\Mw\Omega^2 X=F(t),
\end{equation}
where $\Mw$ is the wall mass, $\tau\propto \alpha^{-1}$ is a damping time, $\Omega$ is the (extrinsic) pinning frequency, and $F(t)$ is a force due to current.
For AC current, $I(t)=I_0 e^{i\omega t}$, where $\omega$ is the frequency, the force is given 
$F(t)=\frac{I(t)}{e}\lt[
\frac{2\hbar S}{\lambda}\beta
-iP\hbar^2\frac{\omega}{\Kp\lambda}\rt]$, 
where $\beta$, given by Eq. (\ref{betadef}), is from momentum transfer and spin relaxation ($\betasf$), and the last term proportional to $\omega$ is from the spin-transfer torque.
The wall under weak current thus shows a forced oscillation of a particle.
By measurering the energy dissipation (from complex resistance), a resonance peak would then appear when $\omega$ is tuned closely around $\Omega$.
From the resonance spectra, the mass and the friction constant were obtained as $\Mw=6.6\times 10^{-23}$kg, $\tau= 1.4\times 10^{-8}$sec.
The experimental result seems to be well described by the rigid-wall picture, and this would be due to a low current density (by factor of $10^{-2}$ compared to DC experiments on metals), resulting in small deformation.  
What is more, from the resonance line shape, the driving mechanism of the domain wall was identified to be the force ($\beta$) rather than the spin-transfer torque.
This finding was surprising at that time, when adiabatic spin-trasnfer torque was considered as the main driving mechanism. 
 The observed force corresponds to the value of $\beta\sim 1.5$, which is too large if $\beta$ arises from spin relaxation $\betasf$ ($\betasf$ is considered to be of the same order as $\alpha$, both arising from spin relaxation). 
If it comes purely from the momentum transfer, the wall resistance is estimated to be 
$R_{DW}=3\times 10^{-4}[\Omega]$, a quite reasonable value. 
 A striking point in this experiment is a significant enhancement of the effect of the force due to resonance, which made possible the low-current operation. 
 On the other hand, the spin-transfer torque is suppressed in the MHz range (as seen from the factor of $\omega$ in the spin-transfer torque term of $F(t)$).

Quite recently, Thomas et al.\cite{Thomas06} succeeded in detecting periodic oscillation of a wall in a confining potential by using ns current pulse at $j=6.9\times10^{11}\Ams$.
The motion was consistent with the rigid wall description in terms of $X$ and $\phiz$.
Periodic variation of chirality, $\phiz$, of a wall was observed in the presence of magnetic field
and current pulse of 10ns at $1\times 10^{12}$A/m$^2$\cite{Hayashi07}.
 The results indicated that the chirality, $\phiz$, plays an important role on the wall propagation, as predicted theoretically \cite{Slonczewski72,TK04,Thiaville05}.

\section{Summary}

We have reviewed theoretical aspects of the current-driven domain wall motion, 
including microscopic derivation of the equation of motion, wall dynamics, and brief discussion on experimental results. 
The effect of current arises from the $s$-$d$ exchange coupling between the local spin and conduction electrons. 
 Treating the non-adiabaticity perturbatively, we have derived fully quantum mechanical expression of torques and forces acting on the wall in terms of Green's functions.
The effect of current on the equation of motion of local spin (modified Landau-Lifshitz-Gilbert equation under electric current) was thus obtained.
Using the results, we derived the equation of motion of the wall.
The wall is assumed to be rigid and one-dimensional, described by two collective coordinates, position $X$ and angle out of easy-plane $\phiz$.
Spin-transfer torque arising from angular momentum conservation was shown to contribute to wall velocity, and  spin relaxation and non-adiabaticity were shown to work as a force on the wall, which induces $\dot\phiz$. 
Solving the equation of motion, we found that there is a threshold current to drive the wall arising from hard axis magnetic anisotropy energy $\Kp$ and/or extrinsic pinning potential $\Vz$. 
Threshold current is determined by $\Kp$ in the intrinsic pinning regime, by $\Vz$, $\Kp$ and force from the current if in extrinsic pinning regime.
Our results would be useful in realizing domain wall motion at low current.

Our formalism can also be applied to describe general spin structures and dynamics under current.
 Extension of our method to first-principle calculation would be useful in realizing fast and efficient switching of magnetization by current.

\acknowledgement
The authors are grateful to Y. Yamaguchi, T. Ono, M. Yamanouchi, H. Ohno, Y. Otani, H. Miyajima, 
M. Kl\"aui, Y. Nakatani, A. Thiaville, E. Saitoh, 
K.-J. Lee, A. Brataas,  R. Egger, M. Thorwart, J. Ieda, 
J. Inoue,  
S. Maekawa 
and H. Fukuyama for valuable discussion.

\appendix


\end{document}